# Heterointerface effects in the electro-intercalation of van der Waals heterostructures

D. Kwabena Bediako,[1,†] Mehdi Rezaee,[2,†] Hyobin Yoo,[1] Daniel T. Larson,[1] Shu Yang Frank Zhao,[1] Takashi Taniguchi,[3] Kenji Watanabe,[3] Tina L. Brower-Thomas,[4] Efthimios Kaxiras[1,5] and Philip Kim[1]*

[1] *Department of Physics, Harvard University, Cambridge, Massachusetts 02138, USA*
[2] *Department of Electrical Engineering, Howard University, Washington, DC 20059, USA*
[3] *National Institute for Materials Science, Namiki 1-1, Tsukuba, Ibaraki 305-0044, Japan*
[4] *Department of Chemical Engineering, Howard University, Washington, DC 20059, USA*
[5] *School of Engineering and Applied Sciences, Harvard University, Cambridge, Massachusetts 02138, USA*

[†] These authors contributed equally to this work

**Molecular-scale manipulation of electronic and ionic charge accumulation in materials is a preeminent challenge, particularly in electrochemical energy storage.[1–4] Layered van der Waals (vdW) crystals represent a diverse family of materials that permit mobile ions to associate reversibly with a host atomic lattice by intercalation into interlamellar gaps.[5,6] Electrochemical ion intercalation in composites of vdW materials is a subject of intense study, motivated principally by the search for high-capacity battery anodes.[7–13] Yet the precise role and ability of heterolayers to modify intercalation reactions remains elusive. Previous studies of vdW hybrids represented ensemble measurements of macroscopic films or powders, which do not permit the isolation and investigation of the chemistry at 2-dimensional (2D) interfaces individually. Here, we demonstrate the electro-intercalation of lithium at the level of individual atomic interfaces of dissimilar vdW layers. Electrochemical devices based on vdW heterostructures[14] comprised of deterministically stacked hexagonal boron nitride (hBN), graphene (G) and molybdenum dichalcogenide ($MoX_2$; X = S, Se) layers are fabricated, enabling the direct resolution of intermediate stages in the intercalation of discrete heterointerfaces and the determination of charge transfer to individual layers. We employ *in situ* magnetoresistance and optical spectroscopy techniques as *operando* probes of reaction progress. Coupled with transmission electron microscopy, low-temperature quantum magneto-oscillation measurements and supported by *ab initio* calculations, our studies at well-defined mesoscopic electrodes show that the creation of intimate vdW heterointerfaces between G and $MoX_2$ engenders over 10-fold accumulation of charge in $MoX_2$ compared to $MoX_2/MoX_2$ homointerfaces, while enforcing a more negative intercalation potential than that of bulk $MoX_2$ by at least 0.5 V. Beyond energy storage, our new combined experimental and computational methodology for manipulating and characterizing the electrochemical behavior of layered systems opens new pathways to control the charge density in 2D electronic/optoelectronic devices.**

The assembly of 2D layers into vdW heterostructures relaxes the requirements on crystallographic commensurability across the vdW interface and enables the creation of atomically precise superlattices[14] with unique interlayer hybrid interfaces that may be synthetically intractable by chemical growth. Owing to the vast structural and electronic diversity of vdW materials, this approach is of great interest for the design of electrodes with optimized interfacial properties for energy storage and electronic devices.[4,7,8] The potential electrochemical opportunities in engineered vdW interfaces include accommodating designed intercalants at superior capacity, increasing intercalation potentials by modifying thermodynamic landscapes, and/or enhancing ion conduction through the interlayer gap by regulating ion diffusion kinetics. Such rational improvements in device performance require a rigorous understanding of the fundamental electrochemical properties (intercalation capacities, equilibrium potentials, kinetics, etc.) of 2D vdW interfaces, which are inaccessible in macroscopic/bulk measurements and often convoluted with side reactions of the electrolyte as well as macroscopic mass and charge transport factors when using conventional ensemble electrochemical methods. Here, the recently developed "Hall potentiometry" (HP) method,[15,16] together with low-temperature quantum transport measurements, directly probes the intercalation process at atomic interfaces to reveal novel synergistic effects of intimate interlayer vdW contact between multiple redox-active 2D layers in a prototypical carbon/metal dichalcogenide system. A complementary potentiostatic technique has elucidated rapid diffusion of Li in graphene bilayers.[16]

To examine the role of the vdW heterointerface in intercalation, we have assembled layers of graphene (G), molybdenum dichalcogenides ($MoX_2$, X = S, Se), and hBN into various precise arrangements. Fig. 1a shows a series of five different heterostructures (Structures **I** through **V**) created using vdW assembly.[17] Structure **I** is a simple vdW structure of graphene encapsulated by hBN, the subject of our previous studies,[15] which serves as a reference point in the present study. The remaining structures (**II-V**) are combinations of atomically thin single crystals of graphene and $MoX_2$ encapsulated by hBN with several vdW heterointerfaces between atomic layers. These electrochemical device architectures are investigated as the working electrodes (WEs) of on-chip micro-electrochemical cells as shown in Fig. 1b and c. Using the HP method (see methods for details), we can monitor both the longitudinal resistance, $R_{xx}$, of the heterostructure WE as well as the Hall carrier density, $n_H$, while the reaction driving force (potential, $E$) is altered. The WEs are encapsulated by electrically inert hBN while exposing the etched boundaries of the vdW stacks for the controlled intercalation reaction. Thus, by measuring $R_{xx}$ and $n_H$ during the electrochemical process with the applied potential $E$ between the WE and the counter electrode, the progress of the electrochemical reaction can be monitored in this mesoscopic system.

Fig. 2 presents an exemplary set of results for electro-intercalation of a heterostructure stack of Structure **II** (hBN/$MoS_2$/G/hBN). From the behaviors of $R_{xx}$ and $n_H$ as a function of $E$, four distinct phases (Phase 1–4) in the electrochemical data can be distinguished corresponding to intermediate stages in the electrochemical reaction of the $MoS_2$/G heterostructure. This in-situ monitoring of $R_{xx}$ and $n_H$ provides more direct information of intercalation staging than that of the traditional electrochemical approach using linear sweep or cyclic voltammetry (see Supplementary Information Fig. 1 for such a comparison), in which the overall electrochemical response is also influenced by any charge transfer reactions at the solid–solution interface of the metal contacts. The transport features in Phase 1 (for $E > -2.3$ V) replicate the purely electrostatic doping behavior observed in electric double layer gating of graphene.[18,19] As we apply an increasingly negative voltage $E$, several intercalation processes occur, revealed by pronounced jumps in $R_{xx}$ and $n_H$. The latter features in Phases 3 and 4, specifically the peak in $R_{xx}$ that occurs in concert with the surge in $n_H$, are key signatures of ion intercalation involving a high mobility graphene layer. The intercalation process engenders a decline in electron mobility as $Li^+$ ions become closely associated with the graphene lattice and act as scattering sites for mobile electrons.[15,16] Ultimately the resistance is driven back down as mounting carrier densities supersede this sudden decrease in mobility.[15]

We find that the deintercalation process (by sweeping $E$ toward 0 V potential) reverses doping and recovers $R_{xx}$ and $n_H$ values similar to those of the pristine heterostructure (Supplementary Information Fig. 1a).

While the transport in the intercalated MoS$_2$/G is largely dominated by conduction through high mobility graphene, as we will discuss later, further insight into the participation of MoS$_2$ in this electrochemical reaction is provided by *operando* photoluminescence (PL) and Raman spectroelectrochemistry. Fig. 2b shows PL data acquired over the course of an $E$ sweep revealing distinct changes to the optical profile of the semiconducting 1$H$- (D$_{3h}$ symmetry) MoX$_2$ layer. Specifically, we find the PL peak consistent with the formation of negatively charged trions (A$^-$)[20] appears at the later stage of intercalation process ($E < -3$ V), signifying strong $n$-doping in MoS$_2$ layer. These data indicate that a highly doped 1$H$-MoS$_2$ phase persists immediately prior to the main intercalation stage, beyond which the PL is fully quenched and Raman (see Supplementary Information Fig. 2a) spectral features of the MoS$_2$ (and graphene) layer are lost due to Pauli blocking.[15,21,22] Deintercalation by reversing the polarization back to 0 V recovers the original Raman spectral features of graphene, yet reveals marked changes in the spectrum of MoS$_2$ as shown in Figure 2c. Most conspicuously, a series of weak, low-wavenumber peaks between ~150 and 230 cm$^{-1}$ emerge. These peaks grow in intensity with increasing number of MoS$_2$ layers—confirming their association with the dichalcogenide—and are still present, albeit slightly diminished, after annealing for 1 h at 300 ºC (Figure 2c, inset). In contrast the Raman peaks for the E$^1_{2g}$ and A$_{1g}$ modes recover spectral intensity after annealing. Figs. 2d–g (and Supplementary Information Figure 2c–g) exhibit the corresponding PL and Raman spectrum homogeneously distributed across the interfacial areas, signifying homogeneity of the intercalation/deintercalation processes.

These spectroscopic results are as expected for an intercalation-induced structural phase transition from the semiconducting $H$ phase to a metallic $T$ (D$_{3d}$ symmetry with octahedral Mo coordination)-type phase with an additional lattice distortion (usually denoted as $T'$), which is favored upon intercalation and/or electron doping of 2$H$-MoS$_2$ beyond ~0.35 Li per MoS$_2$ unit.[23–25] Unlike the $H$ phase, these phases are metallic and therefore the PL is strongly quenched.[26] The low wavenumber Raman features are characteristic of the so-called "J" modes of $T$ and $T'$ phases of MoS$_2$.[26–29] The 154 and 226 cm$^{-1}$ peaks are attributed to the J$_1$ and J$_2$ modes of $T'$-MoS$_2$[27,28] and the 184 cm$^{-1}$ feature is assigned to the J$_1$ mode of $T$-MoS$_2$.[29] The corresponding Raman spectrum peak of the J$_2$ mode for $T$-MoS$_2$ is expected[29] at ~203 cm$^{-1}$ and therefore explains the low wavenumber tail of the $T'$ J$_2$ peak observed in Figure 2c. Notably, we have not observed the emergence of any Raman signatures for lithium polysulfides (746 cm$^{-1}$)[12] during the entire intercalation–deintercalation processes, suggesting the overall chemical integrity of MoS$_2$ remained intact upon lithiation, with a mixed phase of metastable $T$- and $T'$-MoS$_2$ persisting upon deintercalation, and partial recovery of $H$-MoS$_2$ after annealing. We also performed magnetotransport measurements to determine carrier densities induced by intercalation. The total carrier densities for Structure **II** stacks attain values approaching $n_H = 2 \times 10^{14}$ cm$^{-2}$ (Supplementary Information Fig. 3), between three and ten-times the maximal densities observed for intercalated Structure **I** ($2 - 7 \times 10^{13}$ cm$^{-2}$).[15]

Our unique capabilities for *in situ* characterization of electrical properties of electrochemically-modified vdW structures offer a new route to probe the distribution of charge on each 2D layer after intercalation. To uncover these details, we performed low temperature magnetotransport studies in the intercalated vdW heterostructures. Fig. 3a presents the magnetoresistance $R_{xx}$ and Hall resistance $R_{xy}$ for a fully intercalated Structure **II** device at 1.8 K. $R_{xy}$ is linear in magnetic field $B$ from which we estimate $n_H$ equal to $1.0 \times 10^{14}$ cm$^{-2}$. $R_{xx}$ exhibits a pronounced peak near $B = 0$, presumably related to the weak localization behavior due to intervalley scattering of intercalated Li$^+$ ions.[16] As $B$ increases, we observe well-defined Shubnikov-de Haas (SdH) oscillations[30,31] for $B > 3$ T, signifying a high quality 2D electron gas (2DEG) and homogeneity of the lithium-intercalated heterostructure. The periodicity of SdH quantum oscillations with $1/B$ (Fig. 3b) unveils a carrier density $n_{SdH}$ on the order of $2 \times 10^{13}$ cm$^{-2}$, which is five-fold smaller than the total density, $n_H$ estimated from Hall measurement. This discrepancy between $n_{SdH}$ and $n_H$ lies in stark contrast with those observed for Structure **I** (Supplementary Information Fig. 4), and is consistent with a two-channel electronic system, where a higher mobility 2DEG produces SdH oscillations corresponding to a lower density $n_{SdH} < n_H$, while another channel contains the vast majority of electron density ($n_H - n_{SdH}$).

For further quantitative analysis of the electronic band, we measure the SdH behavior at different temperatures, $T$ (Fig. 3b). The decreasing oscillation amplitude with increasing $T$ is linked to the effective mass ($m^*$) of electrons. Fitting SdH data using the standard Lifshitz-Kosevich formalism[30,31] (see Methods), we obtained $m^* = 0.11 m_0$ ($m_0$ is the electron rest mass), close to the value of $0.099 m_0$ that we obtain for intercalated Structure **I** (hBN/graphene/hBN) doped to a density of $\sim 2 \times 10^{13}$ cm$^{-2}$ (additional transport quantities are summarized in Supplementary Information Table 1). We also characterize the low-temperature magnetotransport as a function of back gate voltage $V_g$ applied to the underlying Si substrate (Figs. 3c and d). For Structure **II**, studied above, the graphene monolayer channel is positioned in closer proximity to the back gate, underneath the MoS$_2$ channel (Fig. 3c inset). From the Landau fan diagram, where $R_{xx}$ is plotted as a function of both $V_g$ and $B$ (Fig. 3c), we observe that the SdH quantum oscillations are strongly dependent on $V_g$, pointing to the graphene as the origin of the magneto-oscillations. Were it the case that the MoS$_2$ layer served as the origin of the SdH oscillations, the SdH channel would be electrostatically screened by graphene and the associated density would therefore be independent of $V_g$. Correspondingly, we find that $n_{SdH}$ and $n_H$ exhibit the same dependence on $V_g$ (Fig. 3d), consistent with the bottom-graphene layer (*ca.* $10^{13}$ electrons cm$^{-2}$) serving to shield the overlying MoS$_2$ sheet (*ca.* $10^{14}$ electrons cm$^{-2}$) from the electrostatic influence of $V_g$. In this picture, the dependence of the total density, given by $n_H$, simply follows the dependence of one of its components $n_{SdH}$. We estimate the backgate capacitance, $C = 1.2 \times 10^{-8}$ F cm$^{-2}$ using $\Delta n_H = C \cdot V_g/e$, whose value is in a good agreement with the thickness of SiO$_2$ and hBN layers serving as the gate dielectric.

Having established the carrier density distribution on these heterostructures lies strongly on the metal dichalcogenide layer, we expected that an increase in the number of dichalcogenide layers to two (Structure **III**) would simply furnish carrier densities well in excess of those observed for Structure **II**. Interestingly, we observe that intercalation of Structure **III** stacks ($n_H$ 1.4–1.9 $\times 10^{14}$ cm$^{-2}$, see Supplementary Information Fig. 5) does not lead to carrier densities in excess of typical Structure **II** samples, suggesting that it is the G/MoX$_2$ heterointerface that harbors the vast majority of intercalated ions as opposed to hBN/MoX$_2$ or MoX$_2$/MoX$_2$ interfaces. A significant benefit to our mesoscopic scale device-based approach to electrochemistry and our method of deterministically assembling 2D layers with designed interfaces is the possibility to directly compare electrochemical reactions at different 2D heterointerfaces. By creating heteroarchitectures wherein we design in-plane variations to the structure type along a single graphene monolayer, as depicted in Fig. 4a, we quantitatively examine the dichalcogenide heterolayer effect. Simultaneous measurement of the transport characteristics at different lateral sections of the heterostructure devices during electrochemical polarization (Fig. 4b and see also Supplementary Information Figs. 6 and 7) reveals that the onset of G/MoX$_2$ intercalation takes place at about $\Delta E^o$ = +0.5 to +0.75 V vs. that of G/hBN. Notwithstanding the significantly negative potential, it is noteworthy that the dichalcogenides in these G/MoX$_2$ heterostructures are not decomposed to lithium polychalcogenides as occurs in the bulk (Supplementary Information Fig. 8),[22] indicating a widened window of electrochemical stability. An effect of dimensional confinement on the conversion reaction voltage has been previously observed in MoS$_2$/carbon-nanofiber hybrids.[13] In these systems the outstanding cycling performance is attributed to the restriction of polysulfide/Mo$^0$ nucleation and confinement of product diffusion. We anticipate that a similar mechanism could be pronounced in our 2D vdW heterostructures. Low temperature plots of Hall resistance, $R_{xy}$, as a function of magnetic field (Fig. 4b, bottom left) allow us to precisely identify the total carrier density (and for that matter, lithium ion capacity) in each region of the device. Consequently, these vdW stacks unequivocally demonstrate the critical role of direct graphene–MoX$_2$ heterointerfaces in markedly enhancing the carrier/charge capacities in vdW heterostructure electrodes. We find that sandwiching a graphene monolayer between layers of MoX$_2$ (as in Structures **V** and **V***), thereby creating two graphene–dichalcogenide heterointerfaces produces intercalation capacities more than double those of the "isomeric" Structure **III** region within the same device (Fig. 4b, bottom right). Similar intercalation behaviors were observed in multiple devices (Supplementary Information Fig. 7), confirming the proposition of using intimate vdW contact between different 2D layers as a means of favorably manipulating the equilibrium potentials and capacities in intercalation electrodes. The intercalation onset potentials of the different structures (Figure 4c and Supplementary Information Fig. 9a) emphasize that G/MoX$_2$ interfaces dominate the intercalation properties, since the onset of intercalation is effectively identical across Structures **II** to **V**, and distinctly lying between those of Structure **I** and bulk MoX$_2$. Capacities (Fig. 4c inset and Supplementary Information Figs. 9b,c) as high as $6.2 \times 10^{14}$ cm$^{-2}$ are attainable in Structure **V** devices. However, in all these structures, the graphene density ($n_{SdH}$) exhibits a maximum value of ~$2 \times 10^{13}$ cm$^{-2}$, indicative of a strong preference for charge transfer to the dichalcogenide layers ($n \sim 3 \times 10^{14}$ cm$^{-2}$ each). Assuming additive Li$^+$ capacities, we can estimate the electrochemically accessible capacity of each vdW interface as plotted in Figure 4d, showing the ≥10-fold superiority of the G/MoS$_2$ interface compared to other interfaces. Therefore, the atomic heterointerfaces of vdW materials represent a new material platform to realize engineered functional interfaces for energy conversion and storage. These results highlight the criticality of the graphene heterolayer in enhancing electrochemical charge accumulation in MoX$_2$ while also directing intercalation at a more negative voltage than that of bulk MoX$_2$.

Finally, to provide a comprehensive understanding of the mechanism of intercalation at these 2D interfaces, we explore the atomic scale structural evolution of these layers upon lithiation and delithiation. For this purpose, we fabricate Structure **II** stacks on 50 nm amorphous holey silicon nitride membranes (Figure 5a), and use scanning transmission electron microscopy (STEM) to interrogate these devices in the pristine state before intercalation (Figure 5b) and after one cycle (Fig. 5c) (Additional TEM data from intermediate stages of doping are provided in Supplementary Information Figs. 10 and 11). Although STEM imaging is based on projected atomic structures, we could still obtain atomic resolution images of monolayer $MoS_2$ from the heterostructures by exploiting Z (atomic number)-contrast in high-angle annular dark field (HAADF) STEM imaging and using few-layer hBN crystals. As expected, data from the pristine heterostructure are fully consistent with that of *H*-$MoS_2$ (Fig. 5b). The onset of intercalation results in an increasingly disordered $MoS_2$ lattice evinced by the progressive splitting of the $MoS_2$ Bragg spots in selected area electron diffraction (SAED) patterns. Importantly, even before the peak in $R_{xx}$ we observe this signature of disorder at the edges of the heterostructure, whilst the interior remains pristine (Supplementary Information Figs. 10). Full intercalation results in the observation of a ring in SAED (Figure 5c, inset). Insightfully, aberration-corrected STEM imaging (Figure 5c and Supplementary Information Fig. 11) uncovers crystalline order within domains that are approximately 5–10 nm in size as indicated by fast Fourier transforms (FFT) of the atomic resolution images in specified regions (Figure 5c, right), as well as the inverse FFT shown in Supplementary Information Fig. 11. We also distinctly observe ~1 nm-sized voids in the metal dichalcogenide layer that are reminiscent of what has been reported in previous TEM studies of chemically ($^n$BuLi)-lithiated and exfoliated $MoS_2$.[24] These defects were attributed to beam damage during imaging, which is a possibility for these studies as well owing to the long exposure times required before and during atomic resolution imaging (see Methods). However, this structural disruption is more likely caused by the strain introduced into the $MoX_2$ layer during lithiation and the attendant progression of the *H*- to *T'* phase transformation along the lattice. This structural change has been reported to produce pronounced wrinkles in bulk $MoS_2$ crystals,[22] polycrystallinity in few-layer nanoflakes[32] of $MoS_2$. We often observe visible cracks in the $MoX_2$ layer of our larger devices upon deintercalation (Supplementary Information Fig. 2g). We emphasize, however, that despite these structural defects, the resulting basal plane charge transport in $MoS_2$ layers is reasonably high (as shown in Supplementary Information Table 1), indicating that the intercalation/deintercalation process leaves the $MoS_2$ structure largely intact and as an electrically contiguous layer.

The tuning of intercalation potentials using vdW heterostructures is well explained by the modification of theoretical Li binding energetics as observed in density functional theory (DFT) calculations (Fig. 5d). First, these calculations reveal that the *T'*-$MoS_2$ phase has a considerably stronger binding affinity for Li atoms than *H*-$MoS_2$. Thus, a local phase transformation upon doping should lead to a cooperative effect wherein it becomes increasingly favorable to intercalate Li into that local vdW region as the dichalcogenide undergoes the semiconductor to metal *H*- to *T'*- transformation, thereby lowering the activation barrier for $Li^+$ insertion. This phase transition is manifested by the closing of the band gap and the Fermi level crosses a band with large density of states as shown in Figure 5e and Supplementary Information Fig. 12. Furthermore, since hBN is an inert, wide-gap insulator and non redox-active, the energetics of initial Li intercalation are only slightly perturbed in the case of hBN/*T'*-$MoS_2$ compared to *T'*-$MoS_2$/*T'*-$MoS_2$. In contrast to this situation, graphene (G) heterolayers have a substantially stronger attenuating effect on the binding energy of Li, yet still the reaction is more exergonic than that of Li with hBN/G or G/G[33] (Fig. 5d). As a result, we observe a positive shift in intercalation potential for the G/$MoS_2$ heterolayer compared to hBN/G in Figs. 4b and c.

Taken together, the results of Figures 2–5 are consistent with the electrochemical reaction scheme that is presented in Figure 5f. This mechanism involves charge transfer to both graphene and $MoX_2$ in initial stages of the electrochemical gating process. Dilute concentrations of $Li^+$ ions are intercalated at modest potentials into $MoX_2$/hBN (and $MoX_2$/$MoX_2$) heterointerfaces. However, on the basis of SAED data of our heterostructures and prior observations of sluggish chemical lithiation of bulk $MoS_2$,[22] $Li^+$ ion intercalants of these interfaces appear most concentrated proximate to the heterostructure–electrolyte interface (where the electric field is certainly strongest and some $T'$-$MoS_2$ can be formed locally from electrostatic double-layer gating). The G/$MoX_2$ interface possesses a unique intercalation potential that is more positive than that of G/hBN and as such this is the next interface to undergo intercalation. Eventually a highly doped, 2D nanocrystalline $T'$-$MoX_2$ is formed upon complete intercalation of the graphene–dichalcogenide heterostructure.

We note that typically, in battery electrodes consisting of layered material composites, carbonaceous additives like graphene serve primarily to improve cyclability, particularly over the course of additional conversion reactions that can form insulating and structurally expanded conversion phases.[7,810,12] These approaches do not seek to create or exploit a direct vdW contact between individual atomic layers as a means of tuning the intercalation reaction itself. Our observations for Li-ion intercalation at individual atomic interfaces motivate the use of vdW heterostructuring as a promising strategy in designing new intercalation materials by manipulating the ion storage modes and critical "job-sharing"[4] characteristics of hybrid electrodes. In this regard, our results reveal that the low resistivity of graphene does not necessarily imply that it possesses the majority of electronic charge in the composite material. The vdW heterointerface hosts the $Li^+$ ions, but it is the $MoX_2$ layer that is in fact the most electron-doped. Yet we conclude that the exceptional electronic mobility of graphene (sufficient to display quantum oscillations even after interfacial ion intercalation) furnishes a lower-resistance electronic pathway, notwithstanding a lower partial carrier density, which allows its immediate interface with the $MoX_2$ layer to undergo ionic doping more efficiently than adjacent $MoX_2$/$MoX_2$ homointerfaces. This decoupling of the electron from the ion in an electrochemical scheme is distinct from the case encountered in chemical lithiation reactions where the electron and ion donors are identical (e.g. organolithium reagents). Beyond battery technology, novel ion insertion and ion transport behavior in vdW heterostructures could enable increased control over selectivity in ion separations relevant to detoxification and desalination of water.[34,35] Furthermore, our demonstrated control over intercalation energetics, the resultant spatial carrier density profile, and realization of ultra-high charge densities using vdW heterointerfaces opens up new possibilities for 2D plasmonic device schemes[36] that would require large variations in charge density.

**Acknowledgements** We thank L. Jauregui, I. Fampiou, and G. Kim for important discussions. We thank S. Shirodkar for helpful discussions and sharing data from ref. 29. The major experimental work is supported by the Science and Technology Center for Integrated Quantum Materials, NSF Grant No. DMR-1231319. TEM analysis was supported by Global Research Laboratory Program (2015K1A1A2033332) through the National Research Foundation of Korea (NRF). D.K.B. acknowledges a partial support from the international cooperation project from Korea Institute of Energy Research (KIER). P.K. acknowledges partial support from the Gordon and Betty Moore Foundation's EPiQS Initiative through Grant GBMF4543 and ARO MURI Award No. W911NF14-0247. DFT calculations made use of the Odssey cluster supported by the FAS Division of Science, Research Computing Group at Harvard University; and the Texas Advanced Computing Center (TACC) at the University of Texas at Austin as part of the Extreme Science and Engineering Discovery Environment (XSEDE), which is supported by National Science Foundation grant number ACI-1548562. K.W. and T.T. acknowledge support from the Elemental Strategy Initiative conducted by the MEXT, Japan and JSPS KAKENHI Grant Numbers JP15K21722. Nanofabrication was performed at the Center for Nanoscale Systems at Harvard, supported in part by an NSF NNIN award ECS-00335765.

**Author Contributions** DKB, MR, and HY performed the experiments and analyzed the data. DKB, SYFZ and PK conceived the experiment. DTL and EK performed the theoretical computations. KW and TT provided bulk hBN crystals. DKB, MR, and PK wrote the manuscript. All authors contributed to the overall scientific interpretation and edited the manuscript.

**Author Information** The authors declare no competing financial interests. Correspondence and requests for materials should be addressed to P.K. (e-mail: pkim@physics.harvard.edu).

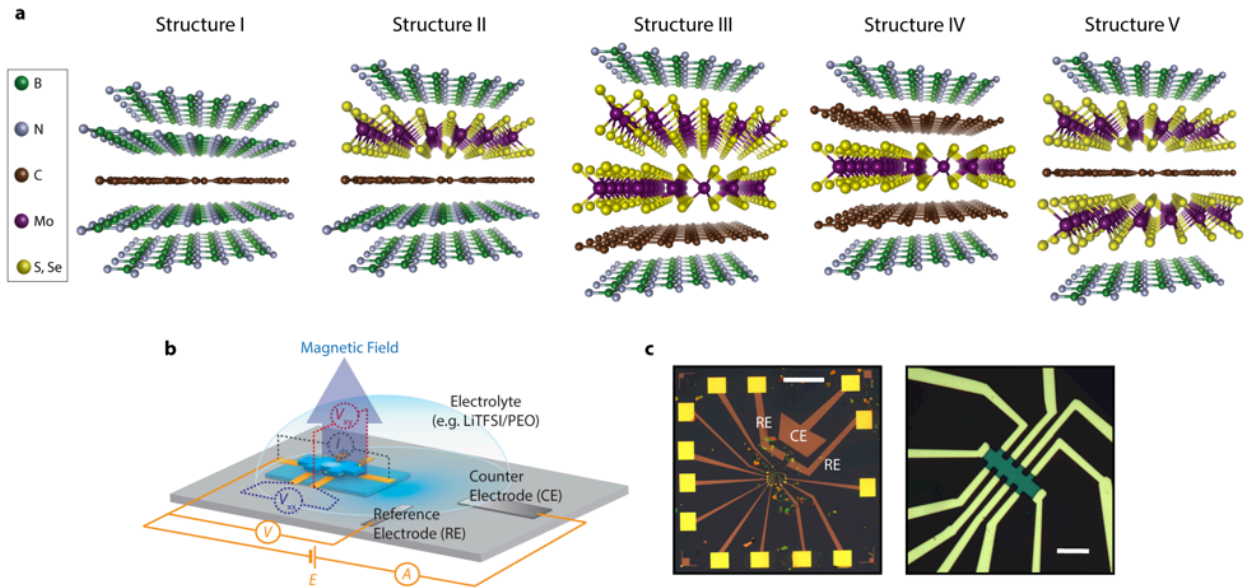

**Figure 1| Van der Waals heterostructures for lithium intercalation. a**, Schematic diagram of the heterostructure series employed for investigation of heterolayers on intercalation capacities and thermodynamics. **b** Schematic model of mesoscopic electrochemical cell where the electrochemical potential, $E$, is applied between the working electrode consisting of a vdW heterostructure and the counter electrode for controlled intercalation. Magnetic fields are applied perpendicular to the heterostructure plane and a small AC current $I_{ds}$ is applied along the heterostructure. Magnetoresistance and Hall resistance $R_{xx}$ and $R_{xy}$ are obtained from the longitudinal and transverse voltages, $V_{xx}$ and $V_{xy}$ by taking their respective ratios to $I_{ds}$. $R_{xx}$ and $R_{xy}$ are measured during the intercalation process by sweeping $E$ in the presence of the electrolyte and the results are compared to the cyclic voltammetry obtained by monitoring the "leakage" current (A). **c**, Optical micrographs of an on-chip electrochemical cell for charge transport and optical measurements during electro-intercalation. Scale bars: left, 500 $\mu$m; right, 10 $\mu$m.

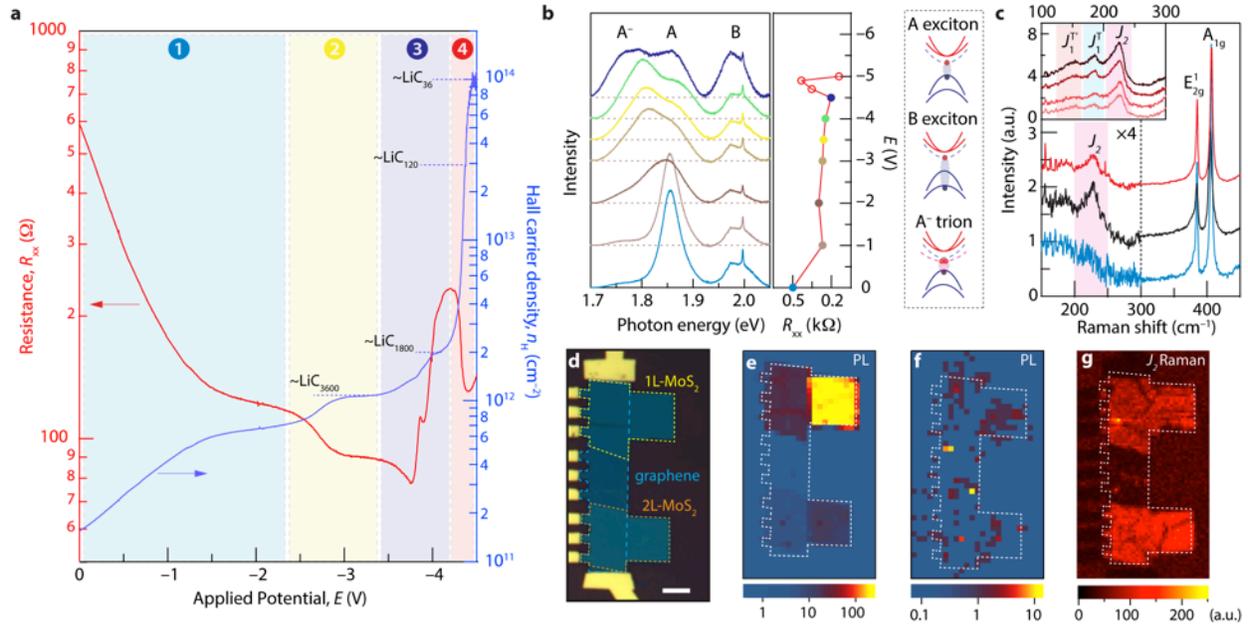

**Figure 2| Intercalation of Structure II graphene–metal dichalcogenide heterostructures. a**, Two-electrode "Hall potentiogram" recorded at 325 K for a graphene–MoSe$_2$ device, showing the change in four-terminal longitudinal resistance, $R_{xx}$ and Hall carrier density, $n_H$ (in the presence of a perpendicular magnetic field of 0.5 T) as a function of applied potential between heterostructure and the counter electrode. Four distinct phases of the intercalation are emphasized along with the approximate stoichiometries of lithium–carbon centers as the sweep progresses. **b**, *Operando* photoluminescence (left) and resistance (middle) measurements acquired at a graphene–MoS$_2$ device over the course of a potential sweep at 325 K. The three pronounced spectral features are assigned to neutral excitons, A and B due to the spin-split valence bands of the dichalcogenide, as well as the formation of negatively charged trions, A$^-$. Spectral baselines are offset (dashed grey lines) to, and color-matched with, their associated resistance–potential data points. **c**, *Ex situ* Raman spectra of a pristine device (bottom), after one cycle to –5 V and back to 0 V (middle), and after annealing for 1 h at 573 K (top). Inset shows Raman spectra after one intercalation–deintercalation cycle and annealing (573 K, 1 h) of hBN-encapsulated (from bottom to top) MoS$_2$, G/MoS$_2$, 2L-MoS$_2$, and G/2L-MoS$_2$. **d**, Optical micrograph of an hBN-encapsulated device consisting of a singular graphene monolayer straddling a monolayer MoS$_2$ flake on one end and a bilayer MoS$_2$ flake at the other as demarcated by the dashed lines. Scale bar: 5 μm. **e–g**, *Ex situ* photoluminescence (**e**,**f**) and Raman (**g**) spatial maps of the device in **d** at room temperature before intercalation (**e**) after intercalation and deintercalation (**f**) and following annealing for 1 h at 573 K (**g**). The Raman map in **g** pertains to the 200–250 cm$^{-1}$ window and therefore represents the spatial intensity of the J$_2$ peak of T/T'-MoS$_2$. The corresponding Raman map for the 350–450 cm$^{-1}$ range (E$^1_{2g}$ and A$_{1g}$ modes) is shown in Supplementary Information Fig. 2g.

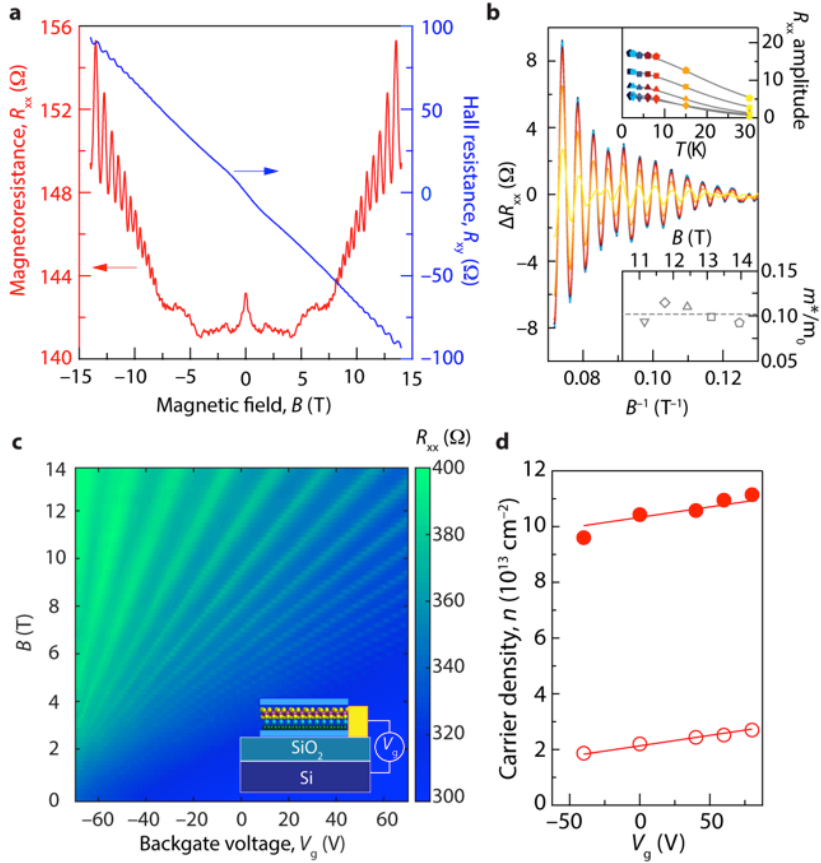

**Figure 3| Quantum transport. a**, Four-terminal magnetoresistance, $R_{xx}$ and Hall resistance, $R_{xy}$ as a function of perpendicular magnetic field strength, $B$, for a Structure **II** graphene–MoS$_2$ device after intercalation at $E = -5$ V. Hall carrier density, $n_H$, is determined to be $1.0 \times 10^{14}$ cm$^{-2}$. **b**, $\Delta R_{xx}$, determined from $R_{xx}$ by subtraction of a polynomial fit to the magnetoresistance background as a function of the reciprocal magnetic field strength, $B^{-1}$, at various temperatures. Shubnikov-de Haas oscillations (SdH) possessing a periodicity of $4.47 \times 10^{-3}$ T$^{-1}$ are observed, consistent with a carrier density, $n_{SdH}$, of $2.2 \times 10^{13}$ cm$^{-2}$. Top inset shows the temperature dependence of SdH amplitude at five values of $B$. Solid lines depict the fit to the data according to the Lifshitz–Kosevich formalism. Bottom inset shows effective masses, $m^*$, extracted from fits at different values of $B$, all around $0.1m_0$ (where $m_0$ is the free electron mas ~$9.11 \times 10^{-31}$ kg). **c,d**, Dependence of charge transport on backgate voltage, $V_g$. **c**, Landau fan diagram $R_{xx}(V_g, B)$ after intercalation, evincing a robust dependence of $n_{SdH}$ on $V_g$, consistent with the underlying graphene layer as the origin of quantum oscillations in these heterostructures. Inset shows schematic of the intercalated heterostructure used with the graphene layer beneath MoS$_2$. Li ions at the hBN/MoS$_2$ and hBN/G interfaces are omitted for clarity. **d**, $V_g$ dependence of $n_{SdH}$ (open circles) and $n_{Hall}$ (filled circles). The lines represent fits assuming a Si backgate capacitance of $1.2 \times 10^{-8}$ F cm$^{-2}$.

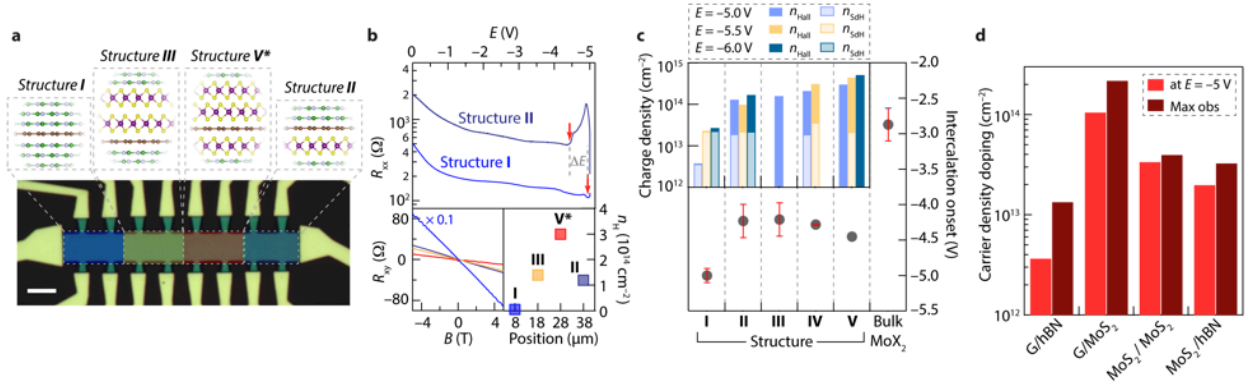

**Figure 4| Tuning intercalation properties with vdW heterolayers. a**, Optical micrograph (false color) of a device consisting of multiple hBN-encapsulated graphene–MoS$_2$ heterostructure types (depicted in the associated illustration) arrayed along a single graphene monolayer. Scale bar: 5 $\mu$m. **b**, Top: Four terminal resistance, $R_{xx}$, as a function of applied potential during electrochemical gating of two regions of the device in **a** demarcated by the assigned contact number. Intercalation (revealed by the sudden rise in resistance indicated at the red arrows proceeds at $\Delta E^o \sim 0.6$ V more positive potentials at the graphene–MoS$_2$ interface than at the graphene–hBN interface, for which intercalation at –5 V is imminent (see also Supplementary Information Figs. 6 and 7). Bottom left: Hall resistances, $R_{xy}$, as a function of applied field, $B$, following polarization of the device in **a** to –5.0 V. Bottom right: Hall carrier densities determined from contacts in each of the four regions of the device. **c**, Onset potentials from intercalation of vdW heterostructures and bulk MoX$_2$ (see also Supplementary Information Fig. 9a) devices. Inset: Mean charge densities (intercalation capacities) achieved after intercalation of devices based on Structures **I**–**V**. Hall carrier densities ($n_H$), indicative of the total density in the heterostructure, are depicted by the overall bar height, whereas the graphene partial carrier densities determined from SdH quantum oscillation data (where available) are indicated by the lighter sub-bars. **d**, Estimated doping level of each interface based on results in **c**.

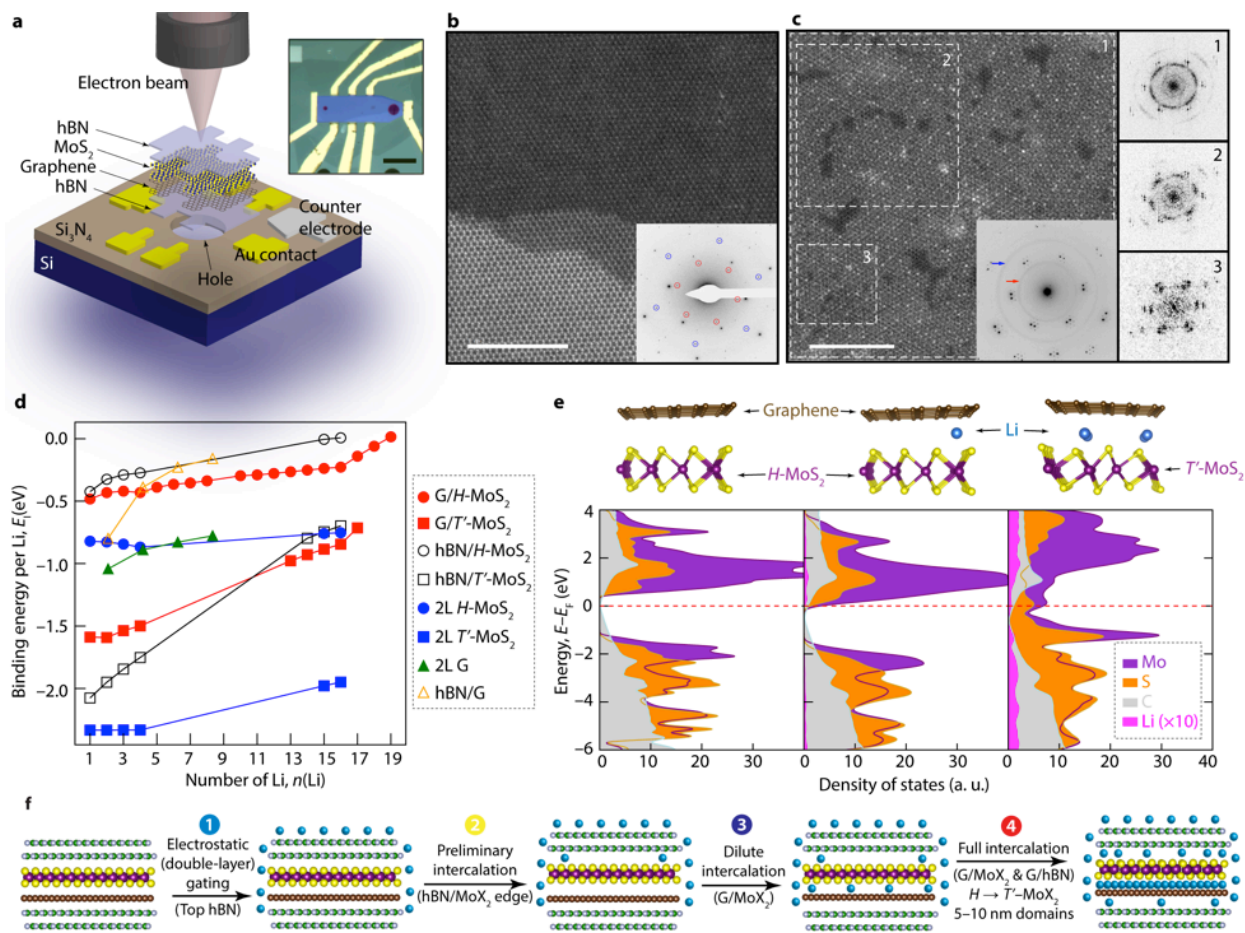

**Figure 5| Structural evolution of vdW heterostructures with electrochemical intercalation. a**, Schematic for vdW heterostructure device assembled onto amorphous holey silicon nitride membranes and subjected to one cycle of intercalation and deintercalation followed by (scanning) transmission electron microscopy, (S)TEM analysis. Inset shows optical microscope image of the one such device. Scale bar 10 μm. **b,c**, High angle annular dark field (HAADF) STEM images of Structure **II** devices before (**b**) intercalation and after one cycle (**c**). Scale bars: 5 nm. Insets show selective area electron diffraction (SAED) patterns using a 300 nm sized aperture. Diffraction features originating from $\{10\bar{1}0\}$ and $\{11\bar{2}0\}$ planes of $MoS_2$ are marked with red and green circles/arrows, respectively. In **c**, fast Fourier transforms (FFT) obtained from the regions indicated with the dashed boxes are shown on the right, showing local crystallinity over ~5 nm domains. **d**, Computed lithium atom binding energy as a function of number of lithium atoms added to a supercell consisting of a layer of 5×5 unit cells of graphene and 4×4 unit cells of $MoS_2$. 2L G and hBN/G data are adapted from ref. 33. **e**, Computed relaxed structures (top) and density of states plots (bottom) for pristine (left), and Li-intercalated (middle: 1 Li; right: 4 Li) heterostructures showing the transition from *H*- (left, middle) to *T'*- (right) $MoS_2$. See Methods for details on DFT calculations. **g**, Mechanism of vdW heterostructure intercalation based on experimental and computational data.

**Methods**

**Sample fabrication.** Samples were fabricated a similar way described in previous work.[1,2] Briefly, mechanical exfoliation of kish graphite (Covalent Materials Corp.) and molybdenum dichalcogenides, $MoX_2$ (X = S, Se) (HQ graphene) onto *p*-doped silicon with 285 nm $SiO_2$ furnishes crystals of the desired thickness, which are identified by optical contrast. Hexagonal boron nitride (hBN) flakes of thickness 15–30 nm are similarly exfoliated and used to pick up graphene and/or $MoX_2$ layers in the desired order. Finally, release of these stacks onto a second flake of hBN results in hBN encapsulated heterostructures that are subjected to annealing in ultra-high vacuum for 30 minutes at 350 ºC. For the devices fabricated on silicon nitride membranes, thinner hBN flakes (≤ 5 nm) were used. Standard electron-beam lithography followed by evaporation of Cr/Pt (1 nm / 9 nm) electrodes is used to define on-chip counter and pseudo-reference electrodes. Reactive ion etching (RIE) using a mixture of $CHF_3$, Ar, and $O_2$ is subsequently used to shape the heterostructure into a hall bar. Another round of lithography is used to delineate an etch mask that overlaps with the protruding legs of the Hall bar. Immediately following another RIE step, the same etch mask is used as the metal deposition mask with Cr/Pd/Au (5 nm / 15 nm /70 nm) contacts. This results in a one-dimensional edge-contact to the active layers and low contact resistances.

**Electrochemical doping & intercalation.** In an Ar-filled glovebox, 3.7 mL of anhydrous (dried with 3 Å molecular sieves) acetonitrile (Sigma-Aldrich) is added to 0.3 g of polyethylene oxide, PEO (Sigma-Aldrich), and 50 mg of lithium bis(trifluoromethane)sulfonamide, LiTFSI. After stirring overnight, a 10–15 µL droplet of this electrolyte solution is cast onto the Si chip possessing the electrically contacted heterostructure stack such that the droplet encompasses both the stack and the counter/reference electrodes. Rapid evaporation of the acetonitrile solvent yields a solid polymer electrolyte for electrochemical studies. Additional extraneous solvent was removed by vacuum drying the electrolyte overnight. Immediately before measurements the measurement device was isolated from ambient moisture and oxygen using a glass cover slip affixed to the chip carrier with vacuum grease. The device is then removed from the glovebox and transferred promptly to the cryostat and vacuum-sealed.

At a temperature of 325 K, the potential between the heterostructure working electrode and Pt counter electrode is swept at a rate of ~1 mV/s in the presence of a small magnetic field, *B*, of 0.5 T. Simultaneously, the resistance of the device is monitored by applying a small AC (17.777 Hz) current ($I_{ds}$) of 0.1–1 µA between the source and drain terminals and measuring the four-terminal longitudinal voltage drop, $V_{xx}$, and Hall voltage, $V_{xy}$, using a lock-in amplifier (Stanford Research SR830). The resistances $R_{xx}$ and $R_{xy}$ are then obtained by $R_{xx} = V_{xx}/I_{ds}$ and $R_{xy} = V_{xy}/I_{ds}$. The Hall carrier density, $n_H$, is then computed from: $n_H = B/(e \cdot R_{xy})$, where $e$ is the elementary charge $1.602 \times 10^{-19}$ C. The Hall mobility, $\mu_H$, during the sweep is also determined from $\mu = (e \cdot n_H \cdot \rho_{xx})^{-1}$, where the resistivity, $\rho_{xx}$ is given by $\rho_{xx} = R_{xx} \cdot W/L$, where $W$ represents the width of the conducting channel and $L$ denotes the length of the channel between contacts. A voltmeter (Agilent 34401A Digital Multimeter) with a high internal impedance of >10 GΩ is used to measure the voltage between the heterostructure working electrode and the Pt pseudo-reference electrode.

Upon reaching the desired potential, the temperature of the system is rapidly cooled to 200 K (10 K/min), thereby freezing the polymer electrolyte and effectively suspending any electrochemical reactions, after which additional magnetic field or temperature dependent sweeps are conducted as desired. Further cooling to base temperature (1.8 K) is carried out at a slower rate of 2 K/min.

Provided potential excursions did not exceed –6 V we found transport behavior to be stable to multiple cycles of these heterostructures.

**Raman & Photoluminescence spectroscopy studies.** Raman and photoluminescence (PL) spectroscopy (Horiba Multiline) is conducted using a 532 nm laser excitation at a power of 5–10 mW with 20 s acquisition times and 4 accumulations. For *operando* studies, the electrochemical cell/device is loaded in a glovebox environment into a cryostat (Cyro Industries of America, Inc.) possessing an optical window. The cryostat is then sealed, transferred out of the glovebox and the measurement chamber evacuated to ultra-high vacuum for spectroelectrochemical measurements. The potential bias is swept at a rate of 2 mV/s to the desired potentials (0, –1, –2, –3, –4, and –5 V) and held at these potentials for acquisition of Raman and Photoluminescence spectra (~10 minutes) before resuming the sweep. After intercalation, the heterostructure is deintercalated by sweeping the potential to +3 V and then back to 0 V. Removal of the electrolyte is accomplished by briefly washing in deionized water followed by isopropanol. Additional spectra are subsequently acquired in this state. The deintercalated heterostructure is then annealed at 300 ºC for 1 h in ultra-high vacuum.

Raman and PL spatial mapping is carried out *ex situ* (after removal of electrolyte) using 1.0 μm step sizes, 2-second acquisition times and 2 accumulations at each pixel/step point.

As noted in the manuscript we do not observe Raman signatures associated with lithium sulfides (743 cm$^{-1}$)[3] during or after intercalation.

---

3. Oakes *et. al.* Interface strain in vertically stacked two-dimensional heterostructured carbon-MoS$_2$ nanosheets controls electrochemical reactivity. *Nat. Commun.* **7**, 11796 (2016).

**Low-temperature charge transport and magnetoresistance analysis**

**Shubnikov-de Haas carrier densities.** Shubnikov-de Haas (SdH) oscillations in $R_{xx}(B)$ arise due to the formation of Landau levels at high magnetic fields.[4] Plotting $R_{xx}(B)$ as a function of $B^{-1}$ confirms that these oscillations are periodic in $B^{-1}$ with a frequency $B_F$. The associated carrier density of the 2DEG, $n_{SdH}$, can then be determined from the relation $n_{SdH} = \left(\frac{g \cdot e \cdot B_F}{h}\right)$ where $g$ is the Landau level degeneracy, $e$ is the elementary charge and $h$ is Planck's constant. For these electron-doped graphene or $MoX_2$ layers it is reasonable to take $g$ as close to 4. Spin–valley locking in the valence band of $H$–$MoX_2$ layers gives rise to degeneracies close to 2, whereas the conduction band-edges are almost spin degenerate leading to degeneracies closer to 4 for electron-doped $H$–$MoS_2$.[5] Theoretical studies to-date do not reveal spin split conduction bands in $T$- or $T'$-phases of $MoS_2$.[6,7] Regardless, the backgate voltage ($V_g$) dependence of Hall and SdH carrier densities provides additional validation for our assignment of the origin of SdH oscillations in the intercalated heterostructures. We find that in the case of a Structure **I** stack consisting of a single graphene monolayer encapsulated by hBN and biased up to $E = -5.5$ V for intercalation, $n_{SdH}$ and $n_H$ equal $\sim 2.6 \times 10^{13}$ cm$^{-2}$ at $V_g = 0$ V, change in concert, and are effectively indistinguishable from each other for $V_g$ between $-100$ V and $+100$ V (Supplementary Information Fig. 7). This reveals SdH and Hall measurements dominated by a single band as expected. In the case of an hBN-encaspulated $MoS_2$–graphene heterostructure (Structure **II**), $n_{SdH}$ changes with $V_g$ in a manner consistent with the capacitance of the $SiO_2$/Si backgate (Manuscript Figures 4d,e). Considering that $n_H$ is the total density of the heterostructure that incorporates $n_{SdH}$, we deduce that the density in only one layer (corresponding to $n_{SdH}$) is dependent on $V_g$. This result reveals that the layer in closest proximity to the backgate (graphene) is responsible for SdH oscillations (lower density), and therefore allows us to determine the degree of charge transfer to the individual $MoX_2$ and graphene layers.

**Transmission electron microscopy.** Aberration-corrected high-angle annular dark field (HAADF) and bright field (BF) scanning transmission electron microscopy (STEM) imaging and selected area electron diffraction (SAED) were conducted by Jeol ARM 200F equipped with cold field emission gun. STEM was operated at 80 kV with the probe convergence angle of 23 mrad. The inner collection semi angle for HAADF STEM imaging was 68 mrad. BF and dark field TEM imaging and SAED were performed on a Tecnai F20 operated at 120 kV. All STEM images shown in Figures 5b and 5c represent the raw, unfiltered data. For the BF STEM image in Supplementary Information Fig. 11d, a Wiener filter[8] was applied to remove noise.

**Effective mass determination, quantum scattering and mobilities.** The effective mass, $m^*$, of the band giving rise to SdH oscillations is determined from the temperature dependence of the SdH amplitude, $\Delta R_{xx}$ (Manuscript Fig. 4b), by fitting these data to the Lifshitz–Kosevich theory,[9]

$$\Delta R_{xx}(B,T) \propto \frac{\frac{\alpha T}{\Delta E_N(B)}}{\sinh\left(\frac{\alpha T}{\Delta E_N(B)}\right)} e^{\left(-\frac{\alpha T_D}{\Delta E_N(B)}\right)}$$

where $B$ is the magnetic field position of the $N$th minimum in $R_{xx}$, $\Delta E_N(B) = heB/2\pi m^*$ is the energy gap between the $N_{th}$ and $(N+1)_{th}$ Landau levels ($m^*$ is the effective mass, $e$ is the elementary charge, and $h$ is the Planck constant), $T_D = \frac{h}{4\pi^2 \tau k_B}$ is the Dingle temperature ($k_B$ is Boltzmann's constant, $\tau_q$ is the quantum lifetime of carriers), and $\alpha = 2\pi^2 k_B$ is the momentum space area including spin degeneracy. In our experiment, $\Delta E_N$ and $T_D$ are the only two fitting parameters. The pre-exponential in this expression is the only temperature dependent portion and permits the straightforward determination of $m^*$ and $\tau_q$. In the case of intercalated Structure **II** (hBN/MoS$_2$/G/hBN), we determine $m^* = 0.11 m_0$, (where $m_0$ is the electron mass) and a Dingle temperature $T_D$ of 36.2 K, which indicates $\tau_q = 33.6$ fs and a mean free path, $l = v_f \cdot \tau_q$ (where $v_f$ is the Fermi velocity that is taken as $10^6$ m s$^{-1}$ for graphene) of ~34 nm. We also determine the quantum mobility, $\mu_q = \frac{e \cdot \tau_q}{m^*} = 558$ cm$^2$ V$^{-1}$ s$^{-1}$ as compared to a Hall mobility $\mu_{Hall}$ of 270 cm$^2$ V$^{-1}$ s$^{-1}$. These values are compared to the parameters obtained for intercalated Structure **I** (hBN/graphene/hBN) in Supplementary Information Table 1.

**Density Functional Theory (DFT) computations.** DFT computations were performed using the projector augemented wave (PAW) method[10] as implemented in the VASP code.[11–14] Van der Waals

---

interactions are included using the zero damping DFT-D3 method of Grimme.[15] The heterobilayer graphene/$MoS_2$ system is modeled with a super- cell consisting of a layer of 5×5 unit cells of fully relaxed graphene, a layer of 4×4 unit cells of $MoS_2$ uniformly compressed by 2.5% (in order to match the graphene lattice spacing), and over 17 Å of vacuum space between successive layers in the direction perpendicular to the layer plane. There are 98 total atoms in the bilayer supercell. All calculations were performed with an energy cutoff of 400 eV. A Γ-centered k-point mesh of 5×5×1 was used for structural relaxations until all forces were smaller in magnitude than 0.01 eV (for 0 and 1 intercalated Li ions) or 0.05 eV (for 2 or more Li ions). The k-point mesh was increased to 11×11×1 for electronic DOS and band structure computations. When relaxing the ions within the supercell, one Mo atom was held fixed as a reference point, and the C atom directly above it was held fixed in the plane of the graphene layer to preserve the registration of the two layers, but was free to relax in the vertical direction. All other atoms were unconstrained. We determine the energetic stability of different intercalation states in various vdW heterostructures by calculating the binding (intercalation) energy per Li atom, $E_I$ (Figure 5d):[16]

$$E_I = \frac{1}{n}[E(M, nLi) - E(M) - nE(Li)]$$

where:

$n$ is the number of Li atoms intercalated

$E(M)$ is the energy of the empty structure M (i.e. 0 Li added),

$E(M, nLi)$ is the energy of the structure M with $n$ number of Li atoms intercalated,

$E(Li)$ is the energy of a Li atom in bulk Li.

*Supplementary Information*

for

**Heterointerface effects in the electro-intercalation of van der Waals heterostructures**

D. Kwabena Bediako,[1†] Mehdi Rezaee,[2†] Hyobin Yoo,[1] Daniel T. Larson,[1] Shu Yang Frank Zhao,[1] Takashi Taniguchi,[3] Kenji Watanabe,[3] Tina L. Brower-Thomas,[4] Efthimios Kaxiras[1,5] and Philip Kim[1]*

[1] *Department of Physics, Harvard University, Cambridge, Massachusetts 02138, USA*
[2] *Department of Electrical Engineering, Howard University, Washington, DC 20059, USA*
[3] *National Institute for Materials Science, Namiki 1-1, Tsukuba, Ibaraki 305-0044, Japan*
[4] *Department of Chemical Engineering, Howard University, Washington, DC 20059, USA*
[5] *School of Engineering and Applied Sciences, Harvard University, Cambridge, Massachusetts 02138, USA*

[†] These authors contributed equally to this work

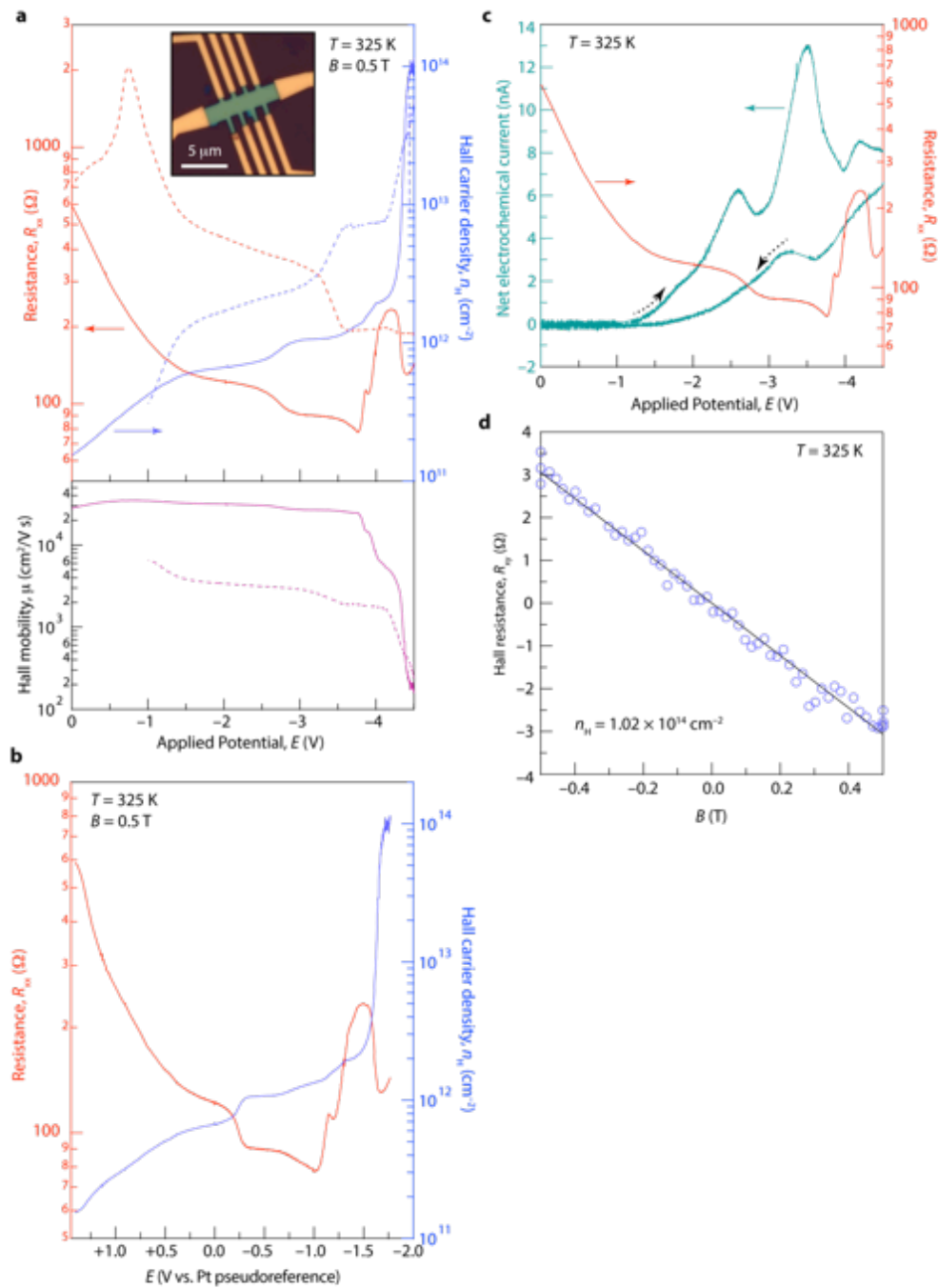

**Supplementary Information Figure 1| Additional electrochemical and Hall data of structure II graphene–MoSe$_2$ stack. a**, Forward (solid lines) and reverse (dashed lines) sweeps of four-probe resistance (red), Hall carrier density (blue), and Hall mobility (purple) as a function of potential at the heterostructure (vs. the counter electrode/electrolyte gate—i.e. in a two-electrode electrochemical configuration) in a LiTFSI/PEO electrolyte at 325 K in the presence of a magnetic field, $B = 0.5$ T. Inset: optical micrograph of heterostructure stack working electrode. **b**, Identical experiment as in **a** with the Resistance (red) and Hall carrier density (blue) plotted as a function of the potential measured relative to a Pt pseudoreference electrode. **c**, Conventional cyclic voltammetric electrochemical current response (gray) overlaid with the resistance (red) over the course of the sweep showing peaks that are difficult to directly assign to any specific reaction, likely incorporating side reactions at the Pt– and Au–electrolyte interfaces. **d**, Hall resistance, $R_{xy}$, as a function of field at 325 K after intercalation ($E = -4.5$ V).

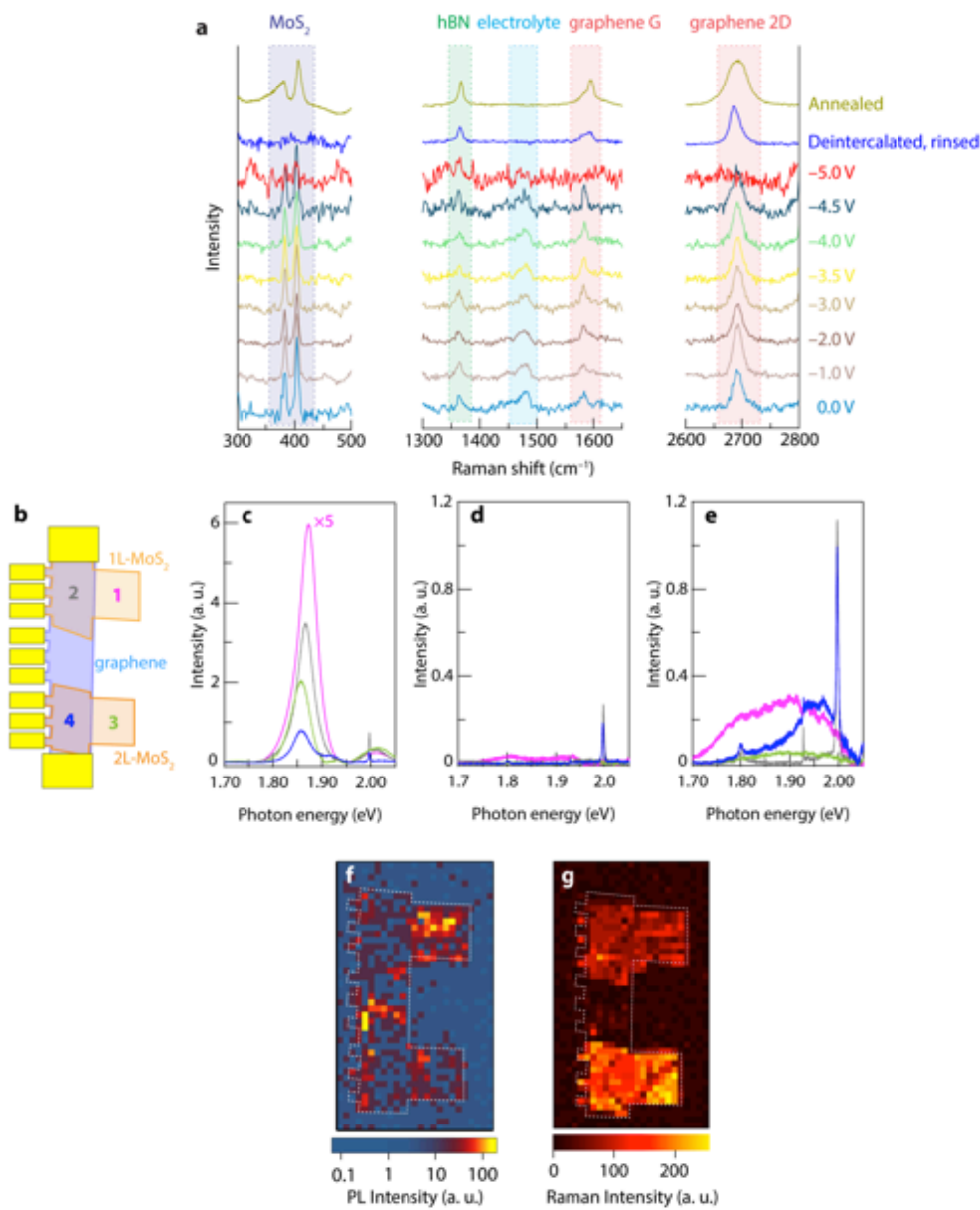

**Supplementary Information Figure 2| Additional Raman & photoluminescence spectroscopy data. a**, Raman spectra of an hBN–graphene–MoS$_2$ Structure **II** device (identical device of Figure 2b in main text) over the course of electrochemical intercalation, showing the disappearance of graphene and MoS$_2$ spectral features after full intercalation at –5.0 V, consistent with Pauli blocking in addition to the $2H \rightarrow 1T'$ phase transition of MoS$_2$. Deintercalation returns graphene peaks, and annealing at 300 ºC for 1 h restores the $2H$–MoS$_2$ peaks. Each spectrum is offset for clarity. **b**–**g**, Schematic diagram (**b**), photoluminescence spectra (**c**–**e**), photoluminescence map (**f**), and Raman map over 350–450 cm$^{-1}$ range (**g**) of an hBN-encapsulated multi-structure device (identical device in Figures 2d–g in main text) consisting of a graphene monolayer straddling a monolayer MoS$_2$ crystal at one end and a bilayer MoS$_2$ crystal at the other. Data were acquired on the pristine stack before intercalation (**c**), after deintercalation followed by removal of electrolyte (**d**), and after subsequent annealing at 300 ºC for 1 h (**e**–**g**). The sharp peak at almost 2 eV is the graphene 2D peak. Photoluminescence spatial maps in the pristine state and after deintercalation are presented in Figure 2e and 2f of the main text and the map of the spatial intensity of the J$_2$ Raman peak of the $T'$ phase (~226 cm$^{-1}$) after annealing is shown in Fig. 2g of the main text.

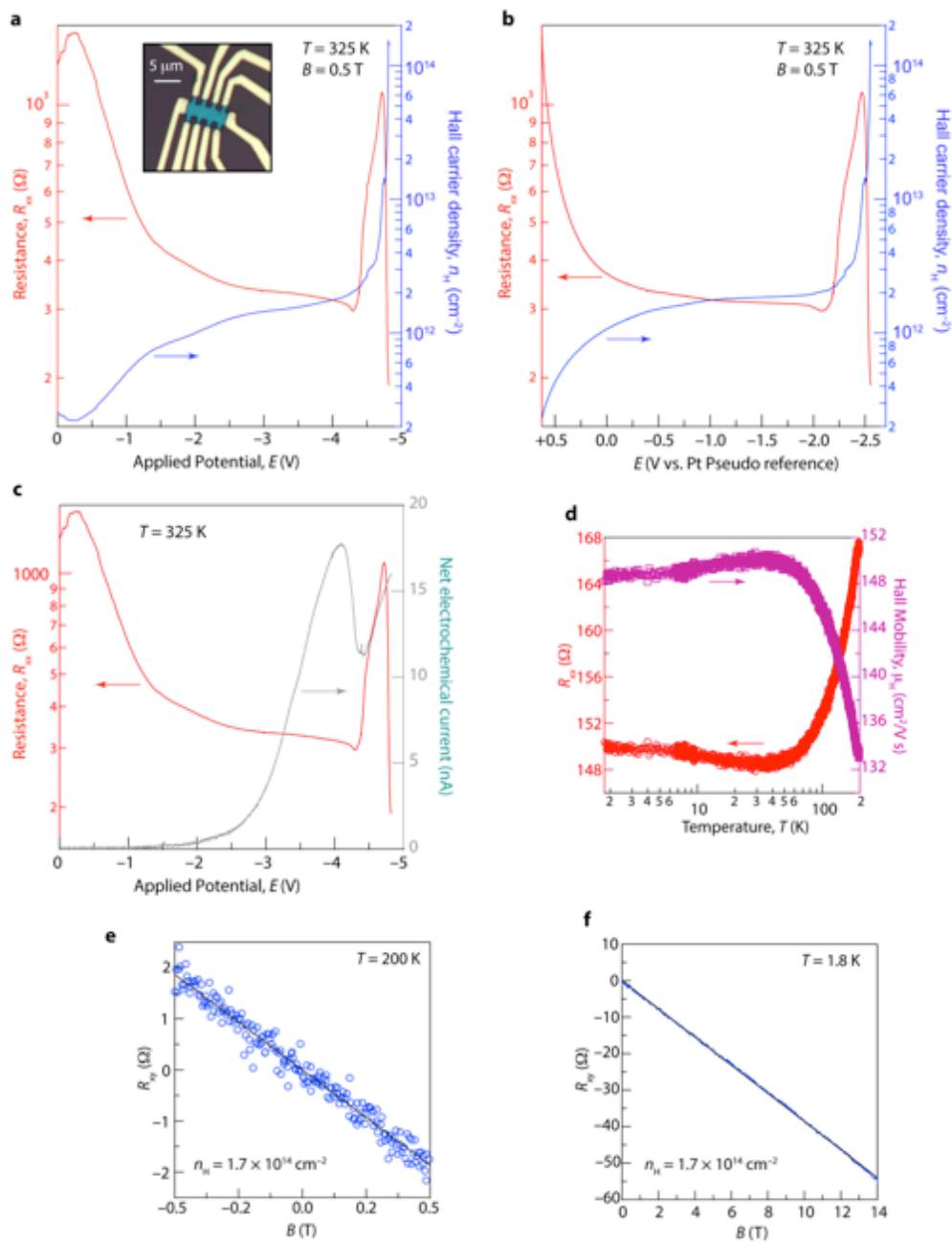

**Supplementary Information Figure 3| Additional electrochemical and Hall data of structure II graphene–MoS₂ stack. a**,**b**, Resistance (red) and Hall carrier density (blue) as a function of potential in a two—potential versus counter—(**a**) and three—potential versus Pt pseudo reference (**b**)—electrode electrochemical configuration in a LiTFSI/PEO electrolyte at 325 K in the presence of a magnetic field, $B$, of 0.5 T. Inset: optical micrograph of heterostructure stack "working electrode". **c**, Conventional cyclic voltammetric electrochemical current response (gray) overlaid with the resistance (red) over the course of the sweep. **d**, Temperature dependence of resistance (red) and Hall mobility (purple) between 200 K and 1.8 K. **e**, Hall resistance, $R_{xy}$, as a function of magnetic field after cooling to 200 K immediately following the termination of a sweep to –4.8 V. **f**, Hall resistance, $R_{xy}$, as a function of magnetic field at 1.8 K.

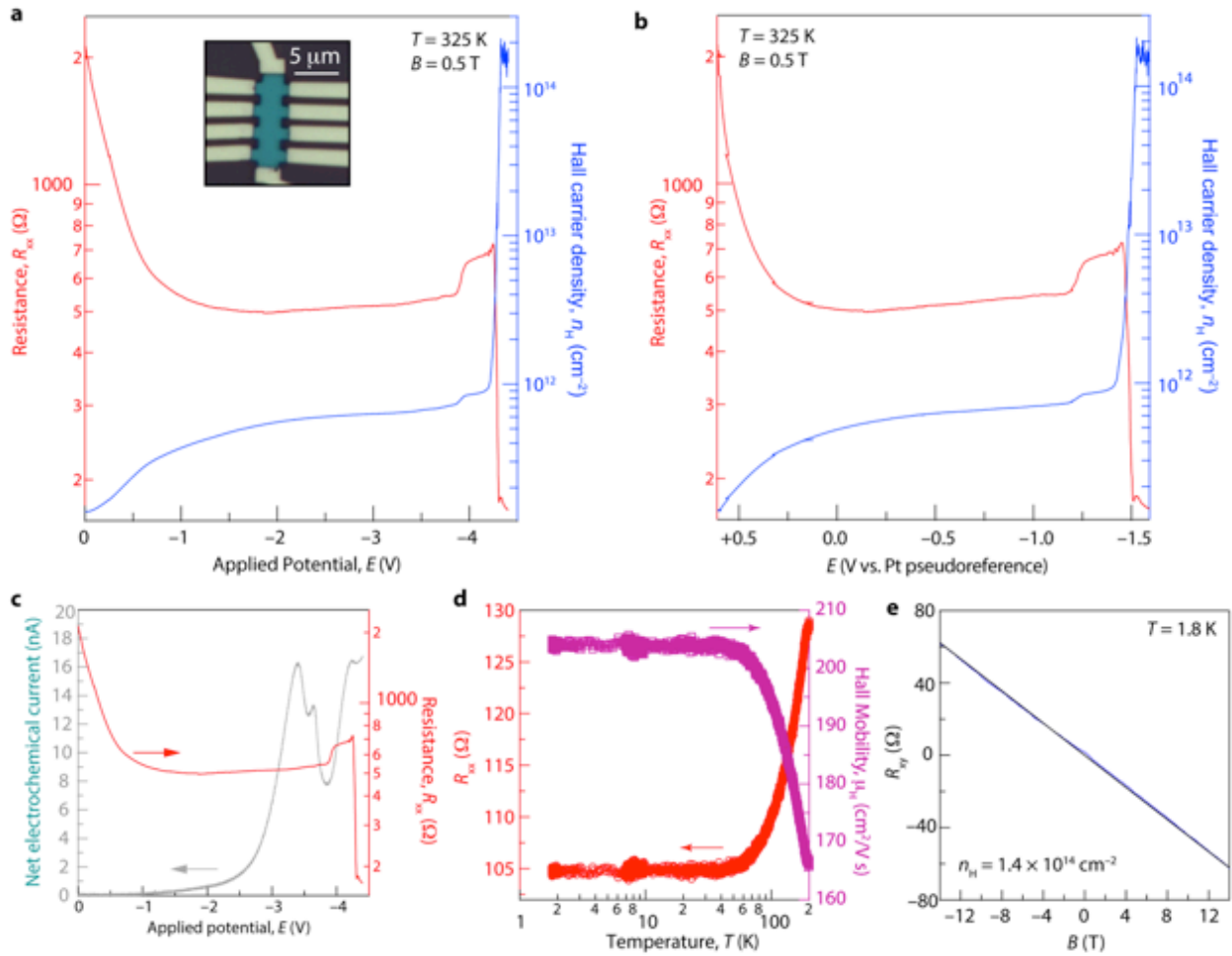

**Supplementary Information Figure 4| Electrochemical and Hall data of structure III graphene–MoS$_2$ stack. a**, Resistance (red) and Hall carrier density (blue) as a function of potential in a two—potential versus counter—(**a**) and three—potential versus Pt pseudo reference (**b**)—electrode electrochemical configuration in a LiTFSI/PEO electrolyte at 325 K in the presence of a magnetic field, $B$, of 0.5 T. Inset: optical micrograph of heterostructure stack "working electrode". **c**, Conventional cyclic voltammetric electrochemical current response (gray) overlaid with the resistance (red) over the course of the sweep. **d**, Temperature dependence of resistance (red) and Hall mobility (purple) between 200 K and 1.8 K **e**, Hall resistance, $R_{xy}$, as a function of magnetic field at 1.8 K. This device shows a carrier density of $1.4 \times 10^{14}$ cm$^{-2}$. Maximum carrier density observed for Structure III devices is $1.9 \times 10^{14}$ cm$^{-2}$.

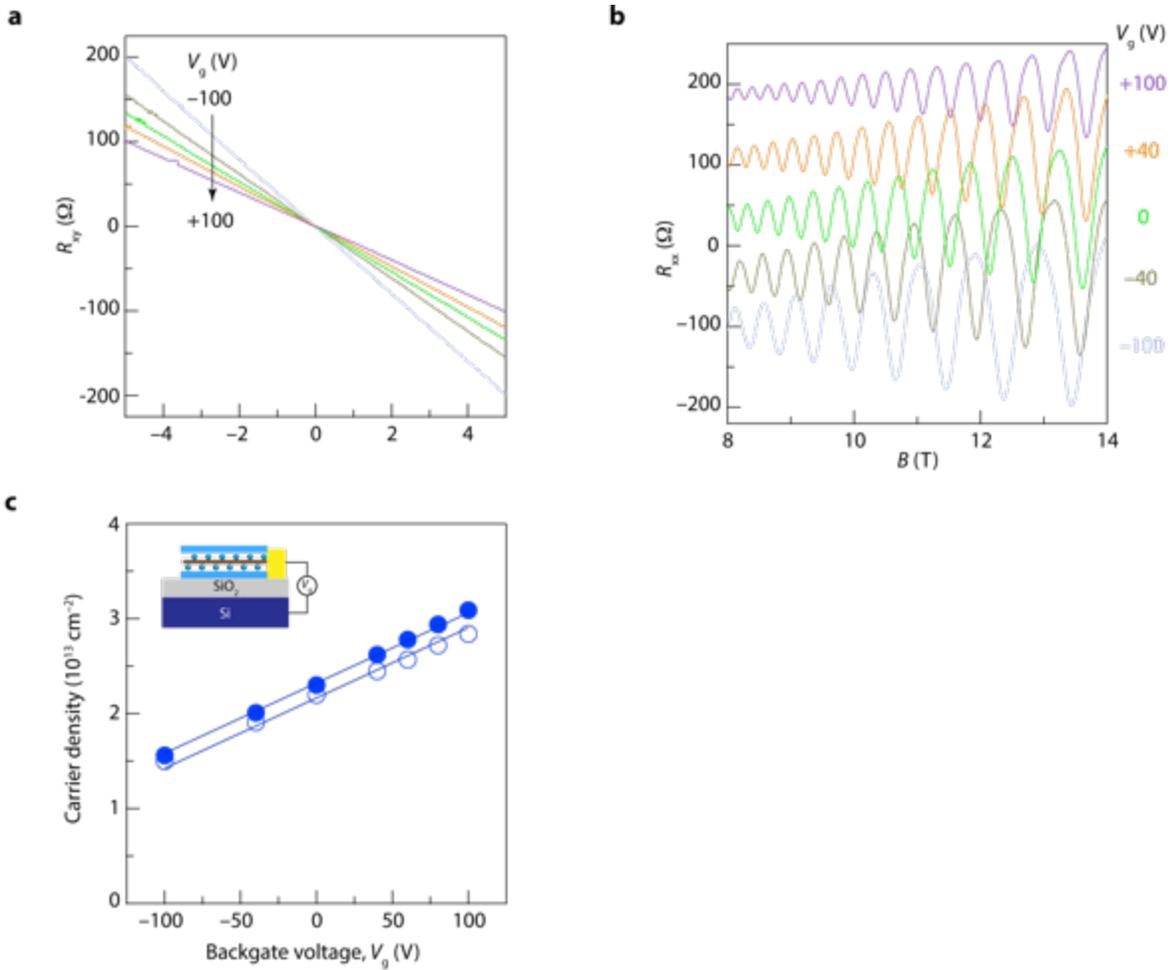

**Supplementary Information Figure 5| Dependence of carrier densities of intercalated heterostructures on backgate voltage. a**, Hall resistance, $R_{xy}$, and **b**, magnetoresistance, $R_{xx}$ (individually offset for clarity), as a function of magnetic field strength, $B$, in the case of a Structure **I** device with varying backgate voltage, $V_g$. **c**, Dependence of change in Hall (filled circles) and SdH (open circles) carrier densities on $V_g$. Solid lines represent fits that assume a Si backgate capacitance of $1.2 \times 10^{-8}$ F cm$^{-2}$.

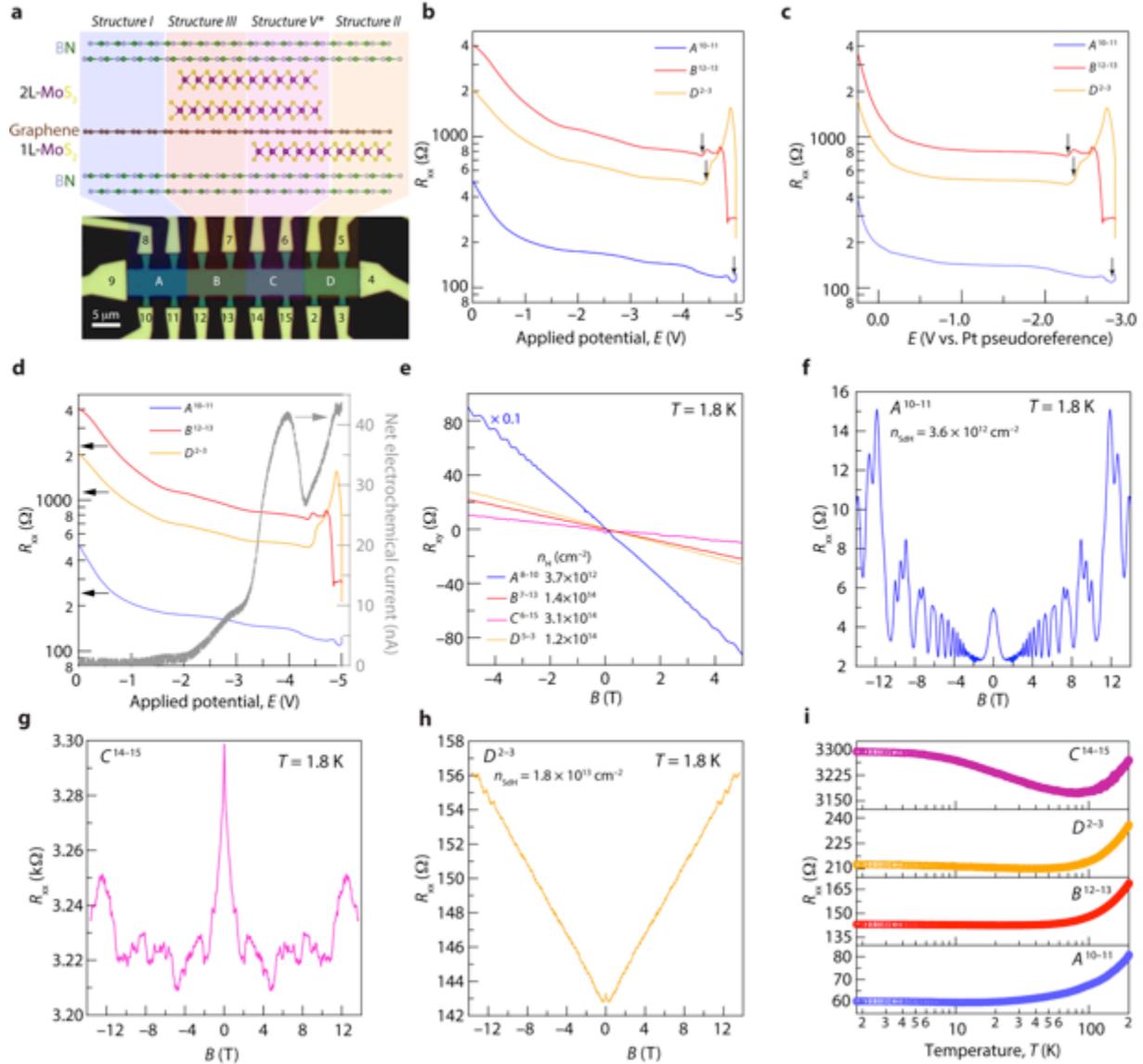

**Supplementary Information Figure 6| Additional data on multi-structure-device #1. a**, Optical micrograph (false color) of a device consisting of multiple hBN-encapsulated graphene–MoS$_2$ heterostructure types (depicted in the associated illustration) arrayed along a single graphene monolayer (identical device as in Fig. 4b of main text). **b,c**, Zonal resistances as a function of potential in a two—potential versus counter—(**b**) and three—potential versus Pt pseudo reference (**c**)—electrode electrochemical configuration in a LiTFSI/PEO electrolyte at 325 K in the presence of a magnetic field, $B$, of 0.5 T. Electrochemical doping of three regions of the device in **a** (demarcated by the assigned contact number) are monitored. Intercalation (indicated by the arrows) initiates at ~0.6 V more positive potentials at zones B and D (structures **III** and **II**) than at A (structure **I**). **d**, Conventional cyclic voltammetric electrochemical current response (gray) of the entire device overlaid with the resistances of the various device regions over the course of the sweep. CV cannot distinguish between the intercalation of G/MoS$_2$ and G/hBN regions in this device. **e**, Hall resistance, $R_{xy}$, as a function of magnetic field at 1.8 K for the different regions of the device after electrochemical polarization up to −5.0 V, displaying the resulting Hall carrier densities obtained. **f–h**, magnetoresistance data at 1.8 K for zones A (**f**), C (**g**), and D (**h**), showing associated SdH carrier densities, $n_{SdH}$ extracted from the periodicities of oscillations in $B^{-1}$. **i**, Temperature dependence of resistance for the various device regions between 200 and 1.8 K during warming.

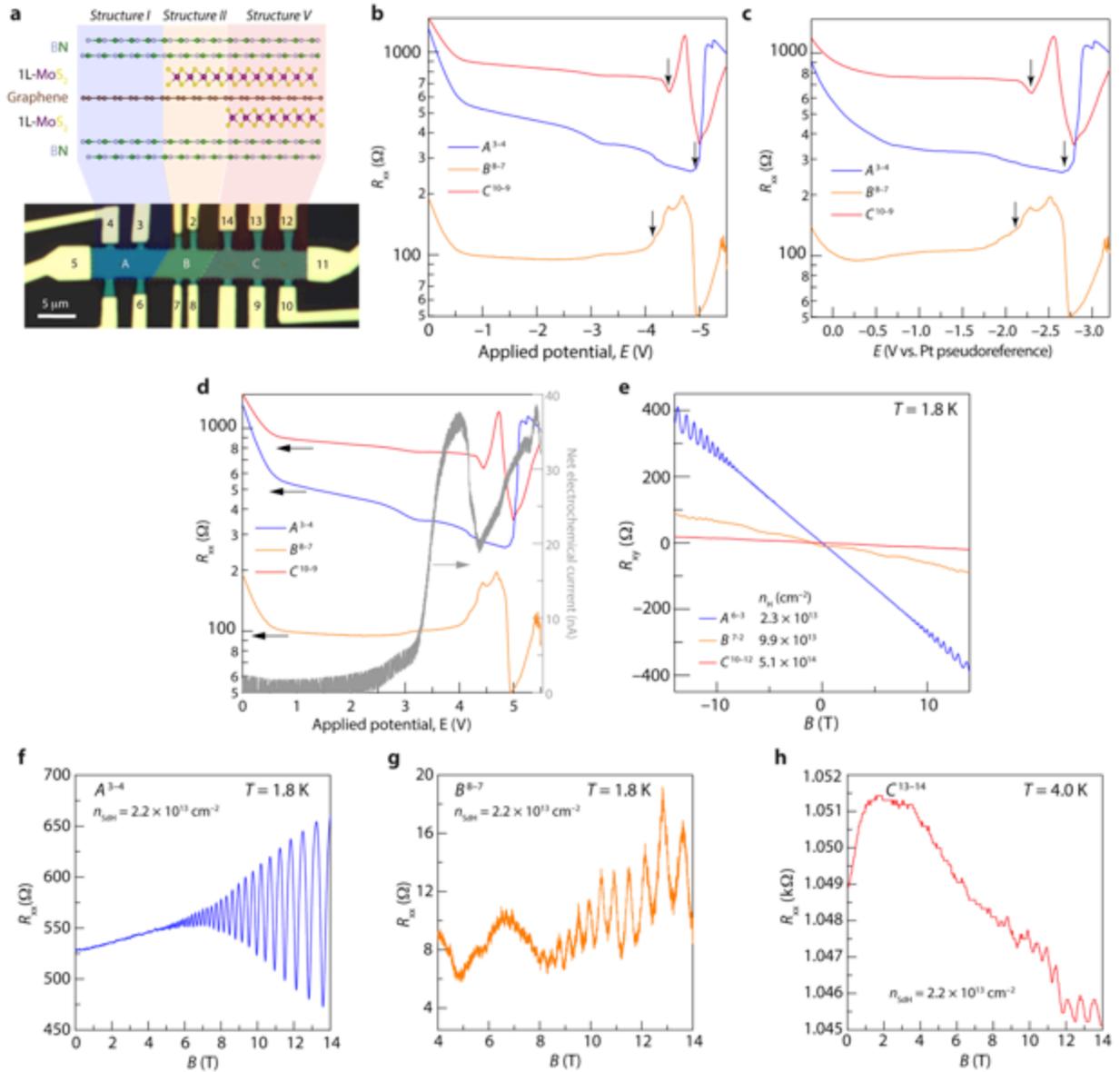

**Supplementary Information Figure 7| Additional data on multi-structure-device #2. a**, Optical micrograph (false color) of a device consisting of multiple hBN-encapsulated graphene–MoS$_2$ heterostructure types (depicted in the associated illustration) arrayed along a single graphene monolayer. **b,c**, Zonal resistances as a function of potential in a two—potential versus counter—(**b**) and three—potential versus Pt pseudo reference (**c**)—electrode electrochemical configuration in a LiTFSI/PEO electrolyte at 325 K in the presence of a magnetic field, *B*, of 0.5 T. Electrochemical doping of the three regions of the device in **a** (demarcated by the assigned contact number) are monitored. Intercalation (indicated by the arrows) initiates at ~0.7 V more positive potentials at zones B (Structure **II**) and C (Structure **V**) than at zone A (Structure **I**). **d**, Conventional cyclic voltammetric electrochemical current response (gray) of the entire device overlaid with the resistances of the various device regions over the course of the sweep. CV cannot distinguish between the intercalation of G/MoS$_2$ and G/hBN regions in this device. **e**, Hall resistance, $R_{xy}$, as a function of magnetic field at 1.8 K for the different regions of the device after electrochemical polarization up to –5.5 V, displaying the resulting Hall carrier densities obtained. **f–h**, Magnetoresistance data at 1.8 K for regions A (**f**), B (**g**), and C (**h**) that reveal associated SdH carrier densities, $n_{SdH}$ from the periodicities of oscillations.

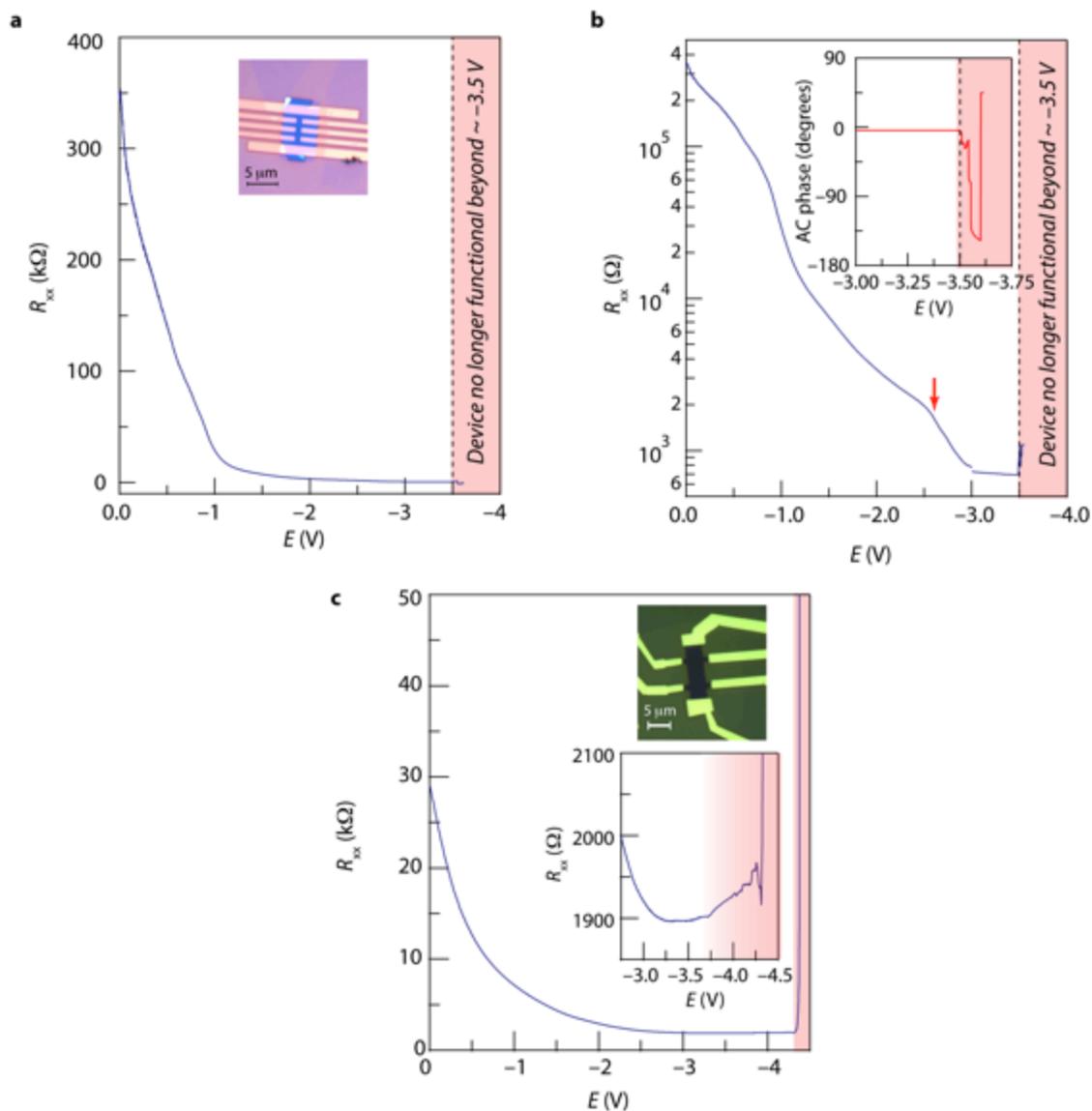

**Supplementary Information Figure 8| Electrochemical gating of non-encapsulated few-layer (4–5 layers) MoX$_2$. a,b**, Four terminal resistance, $R_{xx}$, of a few layer MoSe$_2$ crystal on a linear (**a**) and logarithmic (**b**) scale, during electrochemical gating in an electrolyte comprised of LiTFSI/DEME-TFSI. Intercalation takes place between –2.5 and –3 V (red arrow) and the device loses electrical contact (demonstrated by the disruption in the phase of the lock-in amplifier (inset). **c**, Four-terminal resistance, $R_{xx}$, of a few-layer MoS$_2$ device during electrochemical gating in a LiTFSI/PEO electrolyte. As in **a**, the resistance of this device begins to increase at ~ –3.5 V and is completely insulating beyond –4.25 V, indicative of conversion to lithium polysulfide.

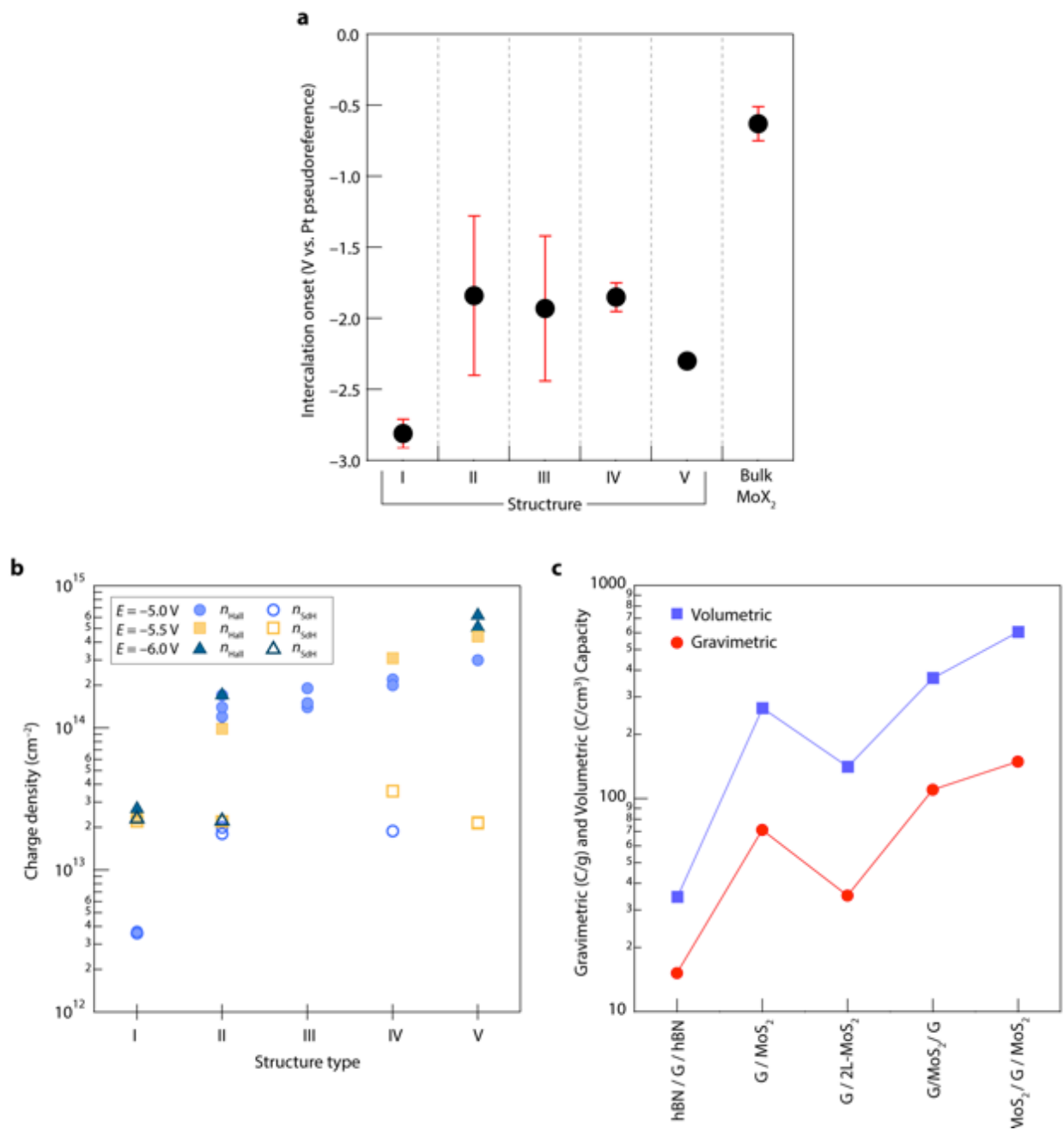

**Supplementary Information Figure 9| Onset potentials and charge capacities of various heterostructures. a**, Intercalation onset potentials (vs. Pt pseudoreference electrode) for different vdW heterostructure types as well as few-layer $MoX_2$. **b**, Carrier densities attained after intercalation of various hBN–graphene–$MoCh_2$ heterostructures. Circles, squares and triangles represent densities reached after intercalation up to −5, −5.5, and −6 V, respectively. Filled symbols designate densities determined from Hall data (revealing approximate $MoX_2$ carrier densities, except in the case of Structure **I**), whereas hollow symbols represent densities extracted from SdH oscillations (revealing graphene carrier densities). **c**, Average capacity values from devices in **b**, expressed in units of C/g (gravimetric capacity) and (C/cm$^3$) volumetric capacity

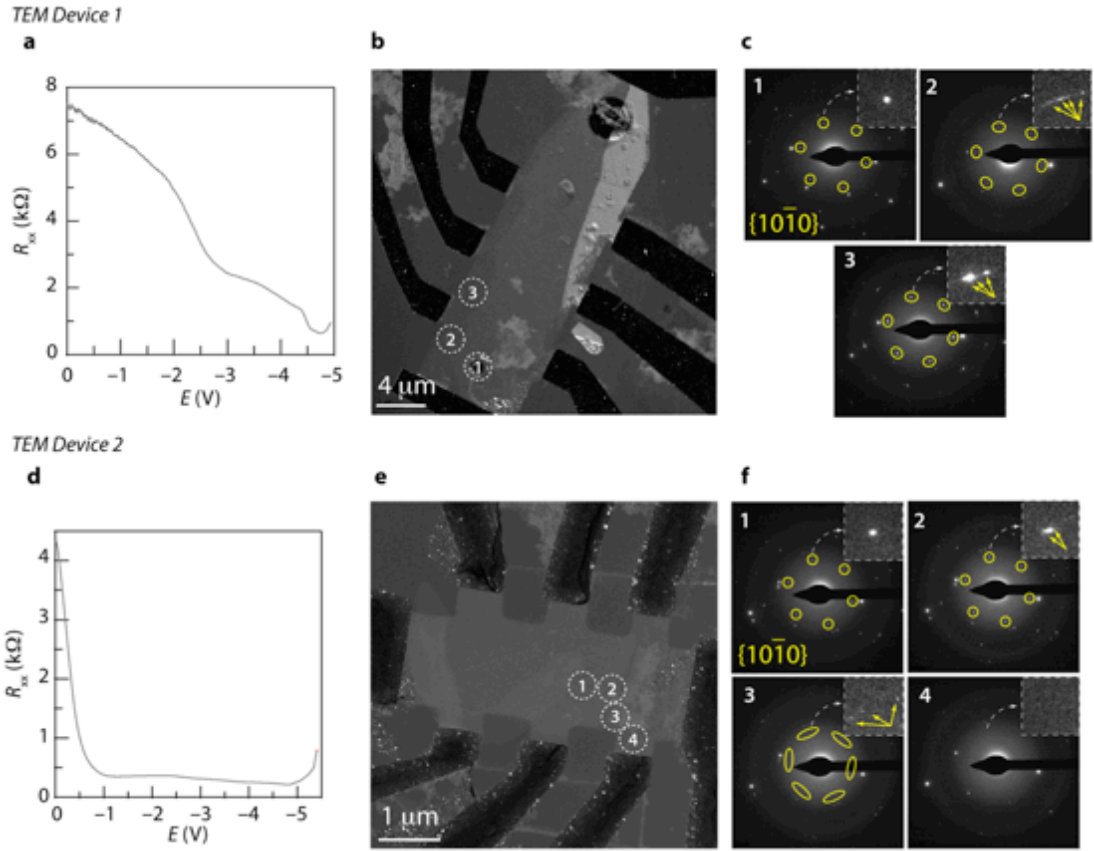

**Supplementary Information Figure 10| Transmission electron microscopy data of incompletely intercalated Structure II devices. a**, Resistance, $R_{xx}$, as a function of applied potential, $E$, of an hBN/MoS$_2$/G vdW heterostructure fabricated onto a 50 nm holey amorphous silicon nitride membrane. The electrochemical reaction is suspended as the upturn in $R_{xx}$ is commencing by immediately sweeping the potential back to 0 V. **b**, $g_{MoS_2} = 11\bar{2}0$ dark-field (DF) TEM image of the device after removal of the electrolyte. **c**, Selected area electron diffraction (SAED) patterns acquired from the regions designated 1, 2, and 3 in **b**. SAED data reveal a pristine MoS$_2$ structure in region 1, but splitting of the Bragg spots (insets) at the edges of the heterostructure (regions 2 and 3) indicative of the formation of two or more domains. **d**, Resistance, $R_{xx}$, as a function of applied potential, $E$, of an hBN/MoS$_2$/G/hBN vdW heterostructure fabricated onto a 50 nm amorphous silicon nitride membrane. The electrochemical reaction is suspended as $R_{xx}$ is approaching a maximum by immediately sweeping the potential back to 0 V. **e**, $g_{MoS_2} = 11\bar{2}0$ DF TEM image of the device after removal of the electrolyte. **f**, SAED patterns acquired using a 300 nm aperture centered on the regions designated 1, 2, 3, and 4 in **e**. SAED data reveal a pristine MoS$_2$ structure in region 1, but strong splitting of the Bragg spots (insets) towards the edge of the heterostructure (region 3) indicative of the formation of multiple domains. In region 4, the diffuse scattering from the underlying amorphous silicon nitride membrane obscures any diffraction features from the MoS$_2$, which in that region must be significantly disordered with any domain sizes $\ll$ 300 nm.

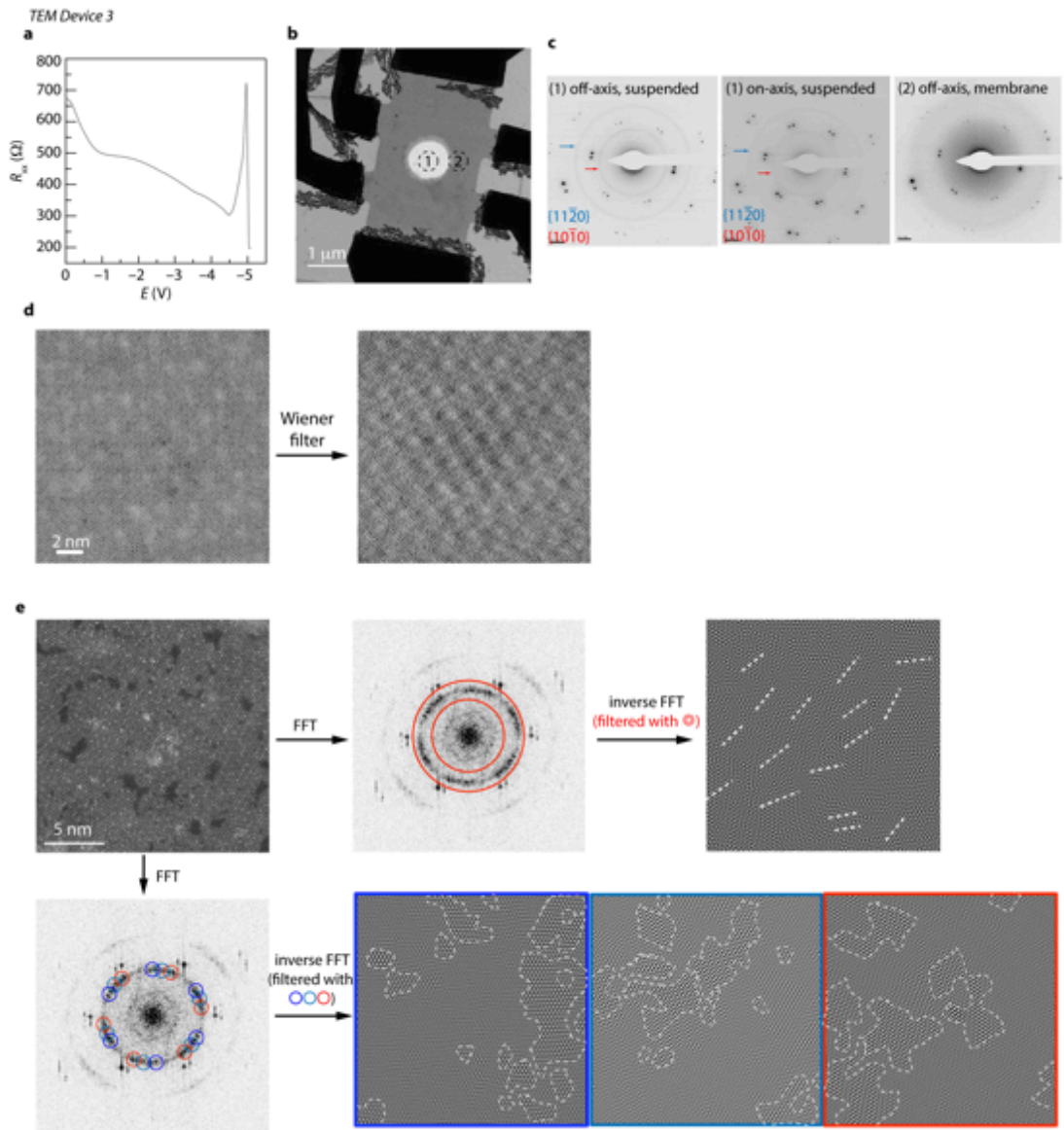

**Supplementary Information Figure 11| (Scanning) transmission electron microscopy, (S)TEM, data of fully intercalated Structure II device. a**, Resistance, $R_{xx}$, as a function of applied potential, $E$, of an hBN/MoS$_2$/G/hBN vdW heterostructure fabricated onto a 50 nm holey silicon nitride membrane. The potential is reversed to 0 V after $R_{xx}$ returns to a minimum (full intercalation) at ~ –5 V. **b**, Bright-field (BF) TEM image of the device after removal of the electrolyte. **c**, Selected area electron diffraction (SAED) patterns acquired using a 300 nm aperture centered on the regions designated 1 and 2 in **b** in both the on-axis (beam perpendicular to the plane of the heterostructure) and off-axis (sample tilted) conditions. The off-axis condition permits the minimization of double diffraction phenomena associated primarily with the top and bottom hBN flakes. SAED data at the suspended (no amorphous silicon nitride) window reveal two rings associated with the MoS$_2$ layer, indicating significant disorder in the x–y plane with a domain size ≪ than the aperture size (300 nm). SAED data acquired over the membrane (region 2) cannot resolve these MoS$_2$ diffraction features owing to the diffuse scattering from the amorphous silicon nitride membrane in that region. **d**, Aberration-corrected BF STEM image of the heterostructure (left: raw data; right: filtered data), which is dominated by the hBN in the structure. The bright periodic patches arise due to the moiré pattern of the two hBN crystals. **e**, Aberration-corrected high angle annular dark field (HAADF) STEM image of the device showing the nanostructure of the MoS$_2$ layer after one cycle. Filtered inverse Fourier transform data resolve x–y rotational disorder in the MoS$_2$ atomic chains (top row white dashed lines), revealing the approximate domain sizes as 5–10 nm (bottom row).

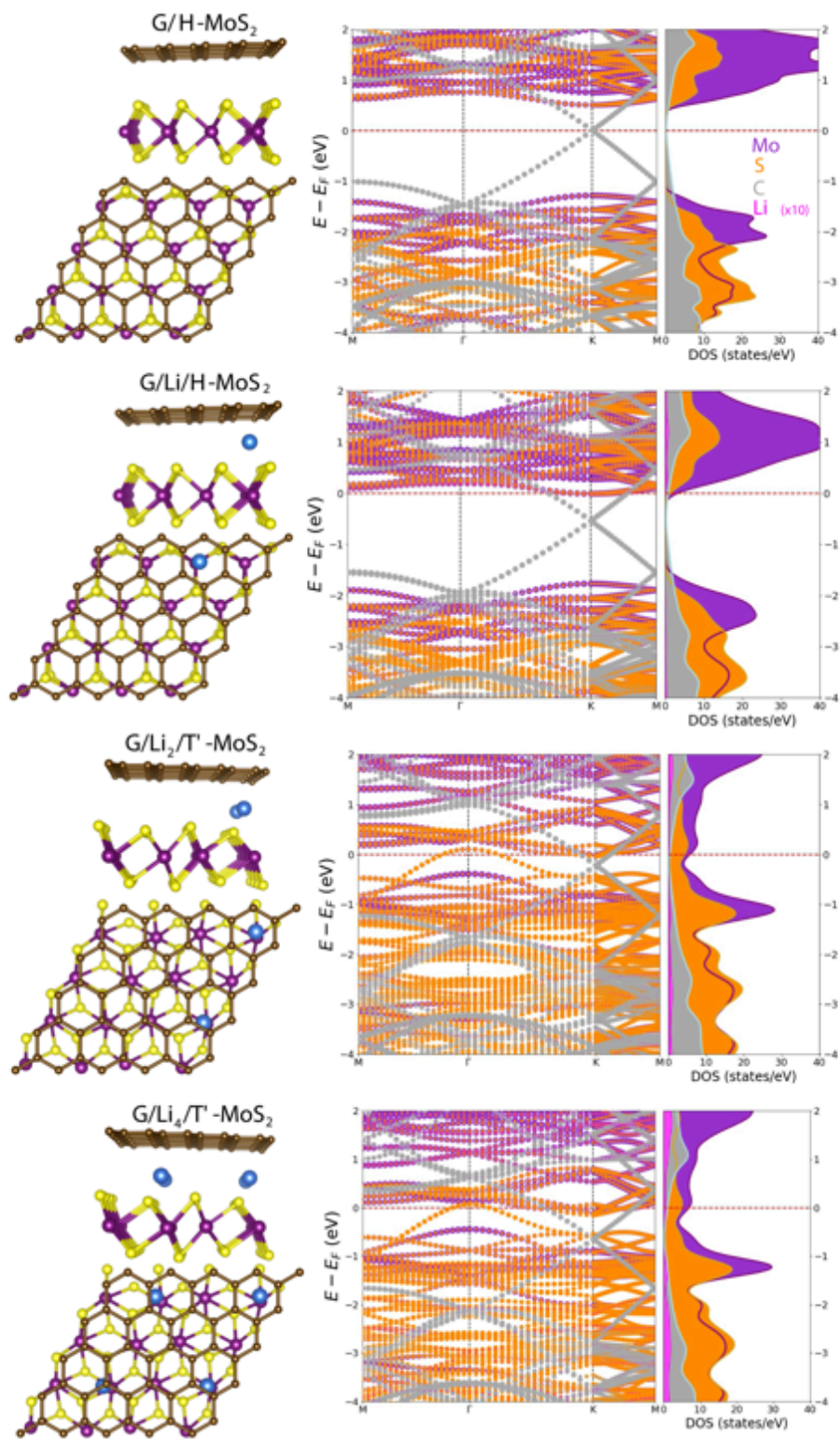

**Supplementary Information Figure 12| Density functional theory (DFT)-computed electronic structures of G/MoS$_2$ heterobilayers over the course of Li intercalation.** Relaxed geometries (left), band structures (middle), and density of states plots (right) for G/MoS$_2$ structures as Li atoms are incrementally added, and the phase of MoS$_2$ is changed from $H$ to $T'$. The reason for the large carrier density in MoS$_2$ versus that in graphene upon intercalation is evident from the relative DOS associated with MoS$_2$ compared to that of G.

| Parameter | Structure I<br>Intercalated ($E$ = –5.5 V)<br>hBN/Graphene/hBN | Structure II<br>Intercalated ($E$ = –5.0 V)<br>hBN/MoS$_2$/graphene/hBN |
|---|---|---|
| $n_\text{H}$ | $2.3 \times 10^{13}$ cm$^{-2}$ | $1 \times 10^{14}$ cm$^{-2}$ |
| $n_\text{SdH}$ | $2.2 \times 10^{13}$ cm$^{-2}$ | $2.0 \times 10^{13}$ cm$^{-2}$ |
| $m^*$ | $0.099 m_0$ | $0.11 m_0$ |
| $T_\text{D}$ | 30.5 K | 36.2 K |
| $\tau_q$ | 39.9 fs | 33.6 fs |
| $l$ | 40 nm | 34 nm |
| $\mu_q$ | 712 cm$^2$ V$^{-1}$ s$^{-1}$ | 557 cm$^2$ V$^{-1}$ s$^{-1}$ |
| $\mu_{Hall}$ | 462 cm$^2$ V$^{-1}$ s$^{-1}$ | 270 cm$^2$ V$^{-1}$ s$^{-1}$ |

**Supplementary Information Table 1| Charge transport parameters.** Comparison of transport parameters for two classes of intercalated heterostructures. The relative similarity in quantum scattering time and mean free compound support for the contention that SdH oscillations observed for intercalated Structure II arise from the graphene sublayer.